\documentclass[aps,prd,preprintnumbers]{revtex4-1}

\usepackage[squaren]{SIunits}
\usepackage{amsmath}
\usepackage{amssymb}
\usepackage{eucal}
\usepackage{graphicx}
\usepackage{subfig}

\begin{document}

\newcommand{\der}{\text{d}}
\newcommand{\mean}[1]{\left\langle#1\right\rangle}
\newcommand{\fulljust}{\setlength{\rightskip}{0pt}\setlength{\leftskip}{0pt}}

\title{Effects of P-wave Annihilation on the Angular Power Spectrum of Extragalactic Gamma-rays from Dark Matter Annihilation}
\author{Sheldon Campbell}
\author{Bhaskar Dutta}
\affiliation{Department of Physics and Astronomy, Texas A\&M University, College Station, Texas 77843, USA}


\preprint{MIFPA-11-22}

\begin{abstract}
We present a formalism for estimating the angular power spectrum of extragalactic gamma-rays produced by dark matter annihilating with any general velocity-dependent cross section. The relevant density and velocity distribution of dark matter is modeled as an ensemble of smooth, universal, rigid, disjoint, spherical halos with distribution and universal properties constrained by simulation data. We apply this formalism to theories of dark matter with p-wave annihilation, for which the relative-velocity-weighted annihilation cross section is $\sigma v=a+bv^2$. We determine that this significantly increases the gamma-ray power if $b/a\agt 10^6$. The effect of p-wave annihilation on the angular power spectrum is very similar for the sample of particle physics models we explored, suggesting that the important effect for a given $b/a$ is largely determined by the cosmic dark matter distribution. If the dark matter relic from strong p-wave theories is thermally produced, the intensities of annihilation gamma-rays are strongly p-wave suppressed, making them difficult to observe. If an angular power spectrum consistent with a strong p-wave were to be observed, it would likely indicate non-thermal production of dark matter in the early Universe.
\end{abstract}

\maketitle

\section{Introduction}
Based on one simple interpretation of astrophysical observations, in the context of $\Lambda$CDM cosmology, it is estimated that about 83\% of the matter in the Universe is dark matter, and that this matter accounts for 23\% of the Universe's total energy content \cite{Komatsu:2008hk, Komatsu:2010fb}. One theory that accounts for the presence of this matter is that of the weakly interacting massive particle (WIMP). In this paradigm, the WIMP is a new stable particle that is produced spontaneously in the early Universe during the Big Bang. WIMP interactions with the Big Bang plasma, for example through WIMP pair production and annihilation, keep its abundance in thermal equilibrium until the Universe becomes too cool to produce new WIMP particles. Annihilation of these particles becomes rare once the rate of expansion of the Universe exceeds the rate of particle annihilation, and the remaining WIMP abundance is said to freeze out. This thermal production of a dark matter relic generates the correct amount of dark matter in our Universe if the WIMP's relative-velocity-weighted annihilation cross section is of the average magnitude $[\overline{\sigma v}]_f\sim\unit{3\times10^{-26}}
{\centi\meter\cubed\per\second}$ at the time of freeze out. If this is the correct theory of dark matter, then we would expect annihilations to be occurring today, predominantly in the densest regions of the Universe. Observation of products from these annihilations not only would give us information about the particle physics nature of the WIMP, but properties of an extragalactic signal also would be rich in information about the large scale structure of matter.

There is an ongoing endeavor to search for signatures of dark matter annihilation in cosmic signals including gamma rays, cosmic rays, and neutrinos. These are looked for: in nearby point sources like the sun, galactic center, and nearby dwarf galaxies; in the diffuse galactic halo; and in the extragalactic distribution \cite{Porter:2011nv}. Indirect signals have already indicated unexpected features. PAMELA \cite{Adriani:2008zr} observes a larger than expected positron fraction in the energy range of $\unit{60-100} {\giga\electronvolt}$, and FGST sees more cosmic electrons than expected at around $\unit{500}{\giga\electronvolt}$ \cite{Abdo:2009zk,*Ackermann:2010ij}. It is possible that these anomalies will be understood in terms of improved models of emission from supernova remnants \cite{Biermann:2009qi}, or pulsar wind nebulae \cite{Profumo:2008ms,*Grasso:2009ma}. Using observations from one indirect signal to constrain these astrophysical models generates predictions for other indirect signals \cite{Kamae:2010ad}. As our understanding of these more standard astrophysical emission processes improves, it becomes more likely that emissions from dark matter annihilation might be extracted. If such a signal is to be identified, precise theoretical predictions of its properties are imperative.

Early estimates of gamma-ray mean intensity and angular power spectrum from extragalactic dark matter annihilation used the spherical halo model of large scale structure \cite{Ando:2005xg,*Ando:2006cr,*Cuoco:2007sh,*SiegalGaskins:2008ge}. The simplest WIMP model was used with $\sigma v\sim\unit{3\times10^{-26}}{\centi\meter\cubed\per\second}$ and a parametrization of the annihilation spectrum motivated from the minimally supersymmetric standard model (MSSM). This formalism was recently generalized to take into account any theory of dark matter annihilation, and a study of different particle physics effects on the mean intensity spectrum of annihilation was presented \cite{Campbell:2010xc}. It is found that in many models the annihilation cross-section is velocity dependent and this has a large impact on the calculation of intensity. Examples of velocity-dependent effects in the annihilation cross section include a p-wave component \cite{Drees:1992am}, Sommerfeld enhancements and resonances \cite{Hisano:2004ds,*ArkaniHamed:2008qn,*Lattanzi:2008qa,*MarchRussell:2008tu}, Breit-Wigner resonances \cite{Ibe:2008ye,*Feldman:2008xs}, and combinations thereof. In this work, we revisit this general formalism, presenting it in a simpler form, and we extend it to the application of calculating the angular power spectrum of the extragalactic annihilation gamma-rays for general velocity dependence of the annihilation cross-section. The present work applies this formulation to the case of p-wave annihilation (the formalism can be applied to the other cases of velocity-dependent annihilation in future work), and offers some preliminary results.

The halo model of large scale structure seems to be an appropriate paradigm for these calculations. Annihilation within smooth halos is dominated in the cores of the halos where the number density is largest. Since halos are predicted by simulations to contain dense substructures, these will also need to be accounted for in order to produce realistic predictions. Current estimates show that the contribution of substructure to extragalactic annihilation within a large halo can increase the signal by a factor on the order of 100, while the galactic signal seen from within the halo is increased by substructure by a factor of only a few \cite{Afshordi:2009hn}. This subhalo effect is not accounted for in this early work and will require attention.

For simplicity, this work assumes that dark matter is distributed throughout the universe in spherical halos. Although halos in general are predicted by simulations to be tri-axial, their cores are nearly spherical. We assume universal radial profiles of the halos' matter density and velocity dispersion, dependent only on the halo's mass and redshift. The velocity distribution is currently approximated to be isotropic (equal radial and transverse velocity dispersion), which is indicated by simulations to be correct deep in the halo cores \cite{Navarro:2008kc}. Where necessary, we assume a locally Maxwell-Boltzmann distribution of the particles, specified by the velocity dispersion at each position\footnote{The assumptions of the velocity distribution may easily be relaxed to more general cases in future work.}. This knowledge is used to determine the average local relative velocity between any two dark matter particles at a particular position. All other needed halo properties, such as concentration, are uniformly taken to be at the ensemble average for the given redshift and halo mass.

For this calculation, it also makes sense to use the rigid halo approximation: far from the halo centers, the dark matter density is low and annihilations are rare, so we may assume the density vanishes beyond some appropriate radius from the halo. Contributions due to overlapping (i.e. merging or unrelaxed) halos are expected to be small and so it is assumed no two halos overlap. In fact, simulations show small halo cores remain relatively intact during mergers with larger halos and contribute as substructures within the parent halo. With all of these considerations, our model of large scale structure is precisely an ensemble of spherical, smooth, disjoint, rigid, mass-universal halos.

A comment about notation: there are two kinds of averages that appear frequently in this paper. For clarity, averages over all space at a fixed time (or over a shell at fixed redshift) are assumed equivalent to a halo ensemble average at that redshift and will be denoted with angle brackets $\mean{\cdot}(z)$. This is in contrast to an average over a distribution at a single position (or over an infinitesimal volume) which will be denoted with an overbar $\overline{\ \cdot\ }(\mathbf{r})$.

This paper is arranged as follows. Section~\ref{sec:formalism} presents the general expressions for calculation of the mean intensity and angular power spectrum of extragalactic gamma-rays from dark matter annihilation. To clarify these results, details leading to the formulae in this section are provided in appendices. The analysis of p-wave annihilation is discussed in Section~\ref{sec:pwave} and results from calculations presented in Section~\ref{sec:results}. We share our conclusions and outlook in Section~\ref{sec:conclusions}. Appendix~\ref{ap:intensity} reviews the intensity of gamma-rays from annihilating dark matter and introduces our notations. The elements of large scale structure that we require are described in Appendix~\ref{ap:lss}, and are applied to the mean intensity and angular power spectrum of the annihilation gamma-ray intensity in Appendix~\ref{ap:signals}. Finally, our evaluations of the angular power spectrum require efficient numerical evaluation of particular Fourier transforms of halo profiles. We share our algorithms for doing so in Appendix~\ref{ap:ft}.

\section{Intensity from position-dependent annihilation cross sections}
\label{sec:formalism}

If we look out with a gamma-ray telescope in direction $\hat{\mathbf{n}}$ along a line-of-sight light shell, with distance and time of photon emission specified by redshift $z$, the intensity of gamma-rays of energy $E_\gamma$ that are due to annihilation of dark matter is (see Appendix~\ref{ap:intensity})
\begin{equation}
  I_\gamma(E_\gamma,\hat{\mathbf{n}})=\int\frac{\der z}{H(z)}W((1+z)E_\gamma,z)
  [\rho^2 \overline{\sigma v}](\hat{\mathbf{n}},z)
\end{equation}
where we define
\begin{equation}
  W(E_\gamma,z)\equiv\frac{1}{8\pi m^2_{\text{DM}}}\frac{1}{(1+z)^3}\frac{\der
    N_\gamma}{\der E_\gamma}(E_\gamma)e^{-\tau(E_\gamma,z)}
\end{equation}
as the intensity window function. In these expressions, $H(z)$ is the Hubble function, $\rho(\hat{\mathbf{n}},z)$ is the dark matter density at the specified position, and $\overline{\sigma v}(\hat{\mathbf{n}},z)$ is the locally-averaged, relative-velocity-weighted, dark matter annihilation cross section at that position. Appearing in the window function is the dark matter particle mass $m_{\text{DM}}$, gamma-ray spectrum per annihilation $\frac{\der N_\gamma}{\der E_\gamma}(E_\gamma)$, and the cosmic opacity to gamma-rays $\tau(E_\gamma,z)$ \cite{Stecker:2005qs,*Stecker:2006eh}.

In s-wave dominated dark matter models where $\overline{\sigma v}$ is a constant, the intensity distribution goes with the square particle number density. However, if the magnitude of the annihilation rate varies over the velocity distribution of cosmic dark matter, then the appropriate position-dependent field that determines the intensity distribution is $\rho^2\overline{\sigma v}$.

Given statistical properties of large scale structure gleaned from simulations in the context of the spherical halo model (described in Appendix~\ref{ap:lss}), we can derive corresponding statistical properties of the extragalactic annihilation gamma-ray intensity. The mean intensity is simply
\begin{equation}
  \mean{I_\gamma}(E_\gamma)=\int\frac{\der z}{H(z)}
  W((1+z)E_\gamma,z)\mean{\rho^2\overline{\sigma v}}\!(z).
\end{equation}
where the mean field is found through an ensemble average over dark matter halos.
\begin{equation}
  \mean{\rho^2\overline{\sigma v}}\!(z)=\int\der^3\mathbf{R}\,\der M
  \frac{\der n}{\der M}(M,z)\ \rho_h^2(R|M,z)\ [\overline{\sigma v}]_h\!(R|M,z)
\end{equation}
Here, the required elements of the halo model are the halo mass function $\frac{\der n}{\der M}(M,z)$, the universal halo density profile $\rho_h(R|M,z)$, and a universal halo profile of the weighted annihilation cross section $[\overline{\sigma v}]_h\!(R|M,z)$ (see Appendix~\ref{ap:lss} for details of these elements).

The angular power spectrum of the intensity signal (detailed in Appendix~\ref{ap:signals}) is
\begin{equation}
  C_\ell(E_\gamma)\approx\frac{1}{\ell^2\mean{I_\gamma}^2\!\!(E_\gamma)}
    \int\frac{\der z}{H(z)}W^2((1+z)E_\gamma,z)k^2
    \overline{P}_{\rho^2\overline{\sigma v}}(k,z)\Big |_{k=\frac{\ell}{r(z)}},
\end{equation}
where the relevant power spectrum of $\rho^2\overline{\sigma v}$ is simply
\begin{eqnarray}
  \nonumber
  \overline{P}_{\rho^2\overline{\sigma v}}(k,z)&=&\int\der M
    \frac{\der n}{\der M}(M,z)\big[\mathcal{FT}\!\{[\rho^2
    \overline{\sigma v}]_h\}\!(k|M,z)\big]^2\\
  &&+\left[\int\der M\frac{\der n}{\der M}(M,z)b(M,z)
    \mathcal{FT}\!\{[\rho^2\overline{\sigma v}]_h\}(k|M,z)\right]^2
    P_{\text{lin}}(k,z).
\end{eqnarray}
Here
\[
  r(z)=\int_0^z\frac{\der\overline{z}}{H(\overline{z})}
\]
is the distance to a position of redshift $z$, $b(M,z)$ is the halo bias function, $\mathcal{FT}\!\{[\rho^2\overline{\sigma v}]_h\}$ is the Fourier transform of the halo profile, and $P_{\text{lin}}(k,z)$ refers to the linear power spectrum at redshift $z$.

The expressions in this section may be applied to any general velocity-dependent annihilation cross section. We will discuss the case of p-wave annihilation now, which commonly appears in various supersymmetric models, for example.

\section{Application to annihilation with p-wave}
\label{sec:pwave}

For s-wave annihilation, $\sigma v=[\sigma v]_0$, a constant. Then the intensity spectrum is simply
\[ \mean{I_\gamma}_0\!(E_\gamma)=[\sigma v]_0\int\frac{\der z}{H(z)}\,W((1+z)E_\gamma
  , z)\,\mean{\rho^2}\!(z)\]
where
\[ \mean{\rho^2}\!(z)=\int\der^3\mathbf{r}\der M\frac{\der n}{\der M}(M,z)
  \rho_h^2(r|M,z), \]
and the angular power spectrum reduces to
\[ C_{0,\ell}(E_\gamma)=\frac{[\sigma v]_0^2}{\ell^2\mean{I_\gamma}_0^2\!(E_\gamma)}
  \int\frac{\der z}{H(z)}W^2((1+z)E_\gamma,z)\,k^2\overline{P}_{\rho^2,\rho^2}
  (k,z)\Big|_{k=\frac{\ell}{r(z)}} \]
with
\begin{eqnarray}
  \nonumber
  k^2\overline{P}_{\rho^2,\rho^2}(k,z)&=&\int\der M\frac{\der n}{\der M}(M,z)
    \big[k\,\mathcal{FT}\{\rho_h^2\}(k|M,z)\big]^2\\
  \nonumber
  &&\quad\quad+\left[\int\der M\frac{\der n}{\der M}(M,z)\left[k\,\mathcal{FT}
    \{\rho_h^2\}(k|M,z)\right]\right]^2 P_\text{lin}(k,z).
\end{eqnarray}
The quantity $k\,\mathcal{FT}\{\rho_h^2\}(k|M,z)$ for the NFW halo profile that we use approaches a constant in the asymptotic $k\rightarrow\infty$ limit (see Appendix~\ref{subap:den2}). Note that, due to the normalization with mean intensity, the angular power spectrum does not depend on the value of the annihilation cross section, $[\sigma v]_0$. In fact, it is a desirable property of the angular power spectrum that it is independent of any uniform constants appearing in the intensity distribution, including constant intensity boost factors that may be associated with halo substructures or non-thermal relic effects, or intensity suppression factors due to p-wave suppression or co-annihilations during freeze out.

For p-wave annihilation, the velocity-weighted annihilation cross section is \[\sigma v = a+bv^2 = [\sigma v]_0\left(1+\frac{b}{a}v^2\right)\] where $[\sigma v]_0=a$ and $b$ are constants, and the cross section halo profile is simply given by Eq.~(\ref{eq:pwavehalo}). In this case, if there is significant dark matter annihilation with square relative velocities $\agt a/b$, then the distribution of produced gamma-rays is coupled to the cosmic dark matter velocity distribution in such a way that regions of high-velocity particles will appear brighter. The intensity spectrum with p-wave annihilation is
\begin{equation}
  \mean{I_\gamma}(E_\gamma)=[\sigma v]_0\int\frac{\der z}{H(z)}W((1+z)E_\gamma,
    z)\mean{\rho^2\left(1+\frac{\lambda b}{a}\sigma_u^2\right)}\!(z)
\end{equation}
where
\begin{eqnarray}
  \mean{\rho^2\left(1+\frac{\lambda b}{a}\sigma_u^2\right)}\!(z)&=&
    \int\der^3\mathbf{r}\der M\frac{\der n}{\der M}(M,z)\rho_h^2(r|M,z)
    \left[1+\frac{\lambda b}{a}\sigma_{uh}^2(r|M,z)\right]\\
  \nonumber
  &=&\mean{\rho^2}\!(z)+\frac{\lambda b}{a}\mean{\rho^2\sigma_u^2}\!(z).
\end{eqnarray}
In these expressions, $\sigma_u^2$ denotes the velocity variance of the dark matter at each position, and $\sigma_{uh}^2(r|M,z)$ is the associated universal halo profile. In this work, we considered local velocity distributions where the mean square relative velocity at each position is related to the velocity variance by $\overline{v^2}=\lambda\sigma_u^2$ for some constant $\lambda$. For Maxwell-Boltzmann distributions, $\lambda=6$.

The effects of the p-wave on the shape of the annihilation spectrum are encoded in the relative contribution of the new second term, due to the p-wave, given by
\begin{equation}
  \label{eq:pwaveeffect}
  \frac{\mean{I_\gamma}\!(E_\gamma|\sigma v=a+bv^2)}
       {\mean{I_\gamma}_0\!(E_\gamma|\sigma v=a)}-1
    =\frac{\lambda b}{a}\Delta_I(E_\gamma)
\end{equation}
with
\begin{equation}
  \label{eq:deltai}
  \Delta_I(E_\gamma)\equiv\frac{
    \int\frac{\der z}{H(z)}W((1+z)E_\gamma,z)\mean{\rho^2\sigma_u^2}\!(z)}
    {\int\frac{\der z}{H(z)}W((1+z)E_\gamma,z)\mean{\rho^2}\!(z)}.
\end{equation}
Other than the dependence on large scale structure in the ensemble averages, $\Delta_I$ depends only on the details of the annihilation spectrum and opacity effects. Note the relative change in intensity diverges for vanishing $[\sigma v]_0$ since the s-wave intensity is zero in this limit.

The angular power spectrum with p-wave annihilations is
\begin{equation}
  C_\ell(E_\gamma)=\frac{[\sigma v]_0^2}{\ell^2\,\mean{I_\gamma}^2\!(E_\gamma)}
    \int\frac{\der z}{H(z)}W^2((1+z)E_\gamma,z)\,k^2
    \overline{P}_{\rho^2\left(1+\frac{\lambda b}{a}\sigma_u^2\right)}(k,z)
    \Big|_{k=\frac{\ell}{r(z)}}
\end{equation}
where the power spectrum is
\begin{eqnarray}
  \overline{P}_{\rho^2\left(1+\frac{\lambda b}{a}\sigma_u^2\right)}(k,z)
    &=&\int\der M\frac{\der n}{\der M}(M,z)\left[\mathcal{FT}\left\{\rho_h^2
    +\frac{\lambda b}{a}\rho_h^2\sigma_{uh}^2\right\}\!(k|M,z)\right]^2\\
  \nonumber
  &&\quad+\left[\int\der M\frac{\der n}{\der M}(M,z)b(M,z)\mathcal{FT}
    \left\{\rho_h^2+\frac{\lambda b}{a}\rho_h^2\sigma_{uh}^2\right\}\!(k|M,z)
    \right]^2 P_{\text{lin}}(k,z)\\
  &=&\overline{P}_{\rho^2,\rho^2}(k,z)+2\frac{\lambda b}{a}
    \overline{P}_{\rho^2,\rho^2\sigma_u^2}(k,z)+\left(\frac{\lambda b}{a}
    \right)^2\overline{P}_{\rho^2\sigma_u^2,\rho^2\sigma_u^2}(k,z).
\end{eqnarray}
For clarification, the mixed power spectrum is
\begin{eqnarray}
  \nonumber
  \overline{P}_{\rho^2,\rho^2\sigma_u^2}(k,z)&=&\mean{\rho^2}\!(z)
    \mean{\rho^2\sigma_u^2}\!(z)P_{\rho^2,\rho^2\sigma_u^2}(k,z)\\
  \nonumber
  &&\!\!\!\!\!\!\!\!\!\!\!\!\!\!\!\!\!\!\!\!\!\!
    =\int\der M\frac{\der n}{\der M}(M,z)\mathcal{FT}\{\rho_h^2\}(k|M,z)
    \mathcal{FT}\{\rho_h^2\sigma_{uh}^2\}(k|M,z)\\
  \nonumber
  &&\!\!\!\!\!\!\!\!\!\!\!\!\!\!\!\!
    +\left[\int\der M\frac{\der n}{\der M}(M,z)b(M,z)\mathcal{FT}\{\rho_h^2\}\!
    (k|M,z)\right]\left[\int\der M\frac{\der n}{\der M}(M,z)b(M,z)\mathcal{FT}
    \{\rho_h^2\sigma_{uh}^2\}\!(k|M,z)\right]P_{\text{lin}}(k,z).
\end{eqnarray}
The biggest challenge in evaluating these expressions is the efficient evaluation of the Fourier transforms. Numerical integration of the Fourier transforms for each integrand sampling during the halo mass and redshift integrations is more time-intensive than is reasonable. See Appendix~\ref{ap:ft} for the efficient algorithms we implemented for evaluation of these transforms for the case of NFW halo profiles.

The relative contribution of the quadratic term in $\sigma v$ to the angular power spectrum is
\begin{equation}
  \label{eq:deltacl}
  \frac{C_\ell(E_\gamma|\sigma v=a+bv^2)}{C_{0,\ell}(E_\gamma|\sigma v=a)}=
  \frac{1+\frac{\lambda b}{a}\Delta^{(1)}_{C_\ell}(E_\gamma)+
    \left(\frac{\lambda b}{a}\right)^2\Delta^{(2)}_{C_\ell}(E_\gamma)}
  {\left[1+\frac{\lambda b}{a}\Delta_I(E_\gamma)\right]^2}
\end{equation}
where each multipole $\ell$ has its own set of power spectrum coefficients
\begin{eqnarray}
  \Delta^{(1)}_{C_\ell}(E_\gamma)&\equiv&\frac{2\int\frac{\der z}{H(z)}
    W^2((1+z)E_\gamma,z)\,k^2\overline{P}_{\rho^2,\rho^2\sigma_u^2}(k,z)
    \Big|_{k=\ell/r(z)}}
  {\int\frac{\der z}{H(z)}W^2((1+z)E_\gamma,z)\,k^2
    \overline{P}_{\rho^2,\rho^2}(k,z)\Big|_{k=\ell/r(z)}},\\
  \Delta^{(2)}_{C_\ell}(E_\gamma)&\equiv&\frac{\int\frac{\der z}{H(z)}
    W^2((1+z)E_\gamma,z)\,k^2\overline{P}_{\rho^2\sigma_u^2,\rho^2\sigma_u^2}
    (k,z)\Big|_{k=\ell/r(z)}}
  {\int\frac{\der z}{H(z)}W^2((1+z)E_\gamma,z)\,k^2
    \overline{P}_{\rho^2,\rho^2}(k,z)\Big|_{k=\ell/r(z)}}.
\end{eqnarray}
It is more convenient to re-express the p-wave effect as
\begin{equation}
  \label{eq:deltaclbar}
  \frac{C_\ell(E_\gamma|\sigma v=a+bv^2)}{C_{0,\ell}(E_\gamma|\sigma v=a)}=
  1+\frac{\frac{\lambda b}{a}\overline{\Delta^{(1)}_{C_\ell}}(E_\gamma)+
    \left(\frac{\lambda b}{a}\right)^2\overline{\Delta^{(2)}_{C_\ell}}(E_\gamma)}
    {\left[1+\frac{\lambda b}{a}\Delta_I(E_\gamma)\right]^2}
\end{equation}
where
\begin{eqnarray}
  \overline{\Delta^{(1)}_{C_\ell}}(E_\gamma)&\equiv&
    \Delta^{(1)}_{C_\ell}(E_\gamma)-2\Delta_I(E_\gamma),\\
  \overline{\Delta^{(2)}_{C_\ell}}(E_\gamma)&\equiv&
    \Delta^{(2)}_{C_\ell}(E_\gamma)-\Delta_I^2(E_\gamma).
\end{eqnarray}
It is interesting to note that this has a well-defined finite limit in the vanishing $a$ limit, and that $\overline{\Delta^{(1)}_{C_\ell}}$ does not contribute in that limit.

\section{Sample Calculations}
\label{sec:results}

We calculated the angular power spectrum for particle physics models with p-wave annihilation components that were used in \cite{Campbell:2010xc}. We find the results to be nearly universal against the different particle physics phenomenologies, suggesting that the effect of p-wave annihilation on the angular power spectrum is largely determined by details of large scale structure. We will present results of the calculation for the most typical model. The angular power spectra results for the other models were within $50\%$.

The typical particle physics model is taken from minimal supergravity (mSUGRA), with parameters $m_0=\unit{2569}{\giga\electronvolt}$, $m_{1/2}=\unit{395}{\giga\electronvolt}$, $\tan\beta=10$, $A_0=0$, and $\mu>0$. The dark matter is the lightest neutralino $\tilde{\chi}_1^0$ with mass $m_{\tilde{\chi}_1^0}=\unit{150}{\giga\electronvolt}$. The needed particle physics details of this model were calculated using DarkSUSY~5.0.5 \cite{Gondolo:2004sc}, interfaced with ISAJET~7.78 \cite{Paige:2003mg} and FeynHiggs~2.6.5.1 \cite{Hahn:2009zz}. This model was calculated to have a thermal relic density of $\Omega_{\text{DM}}h^2=0.114$. This chosen example is from the focus point region of parameter space.

\begin{figure*}
  \subfloat{\includegraphics[width=0.4\textwidth]
    {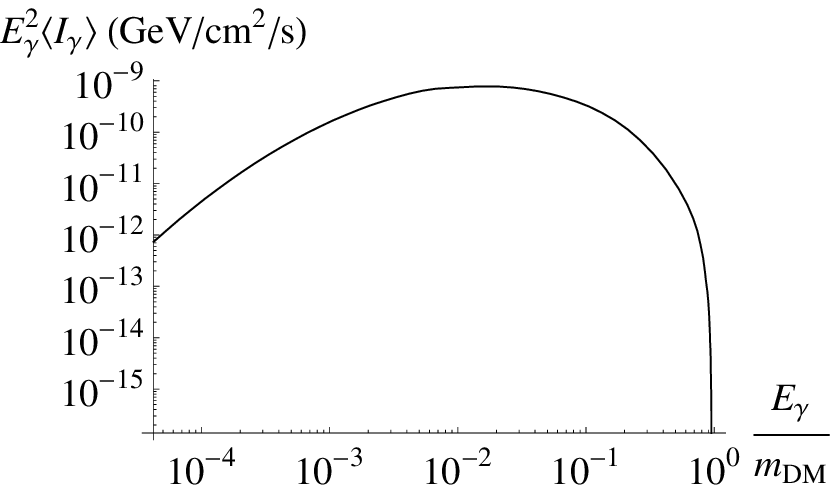}}\quad
  \subfloat{\includegraphics[width=0.4\textwidth]
    {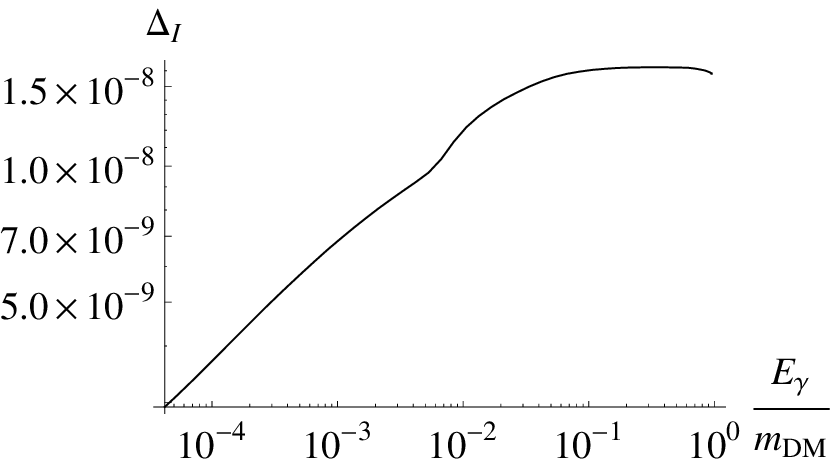}}
  \caption{\label{fig:intensities}\fulljust Left: Intensity spectrum of diffuse gamma-rays from extragalactic annihilating dark matter for the particle model described in the text and for matter distributed according to the spherical halo model, and assuming a thermal dark matter relic. Right: The relative effect of p-wave annihilation on the spectral shape for this model, per $\lambda b/a$. Its magnitude of $10^{-8}$ is typical for any particle physics model. \hfill\mbox{}}
\end{figure*}

In Figure~\ref{fig:intensities}, we plot the mean intensity spectrum of the diffuse extragalactic annihilation gamma-rays, and the relative effect of the p-wave on the spectral shape. If we had chosen a model in the stau neutralino coannihilation region of parameter space at low $\tan\beta$, where the s-wave of the dark matter annihilation cross section is strongly helicity suppressed making the p-wave component strong, we would find that the intensity curve is similarly shaped but is as suppressed as the cross-section's s-wave. However, $\Delta_I$ is very similar between the different models.

\begin{figure*}
  \subfloat{\includegraphics[width=0.4\textwidth]
    {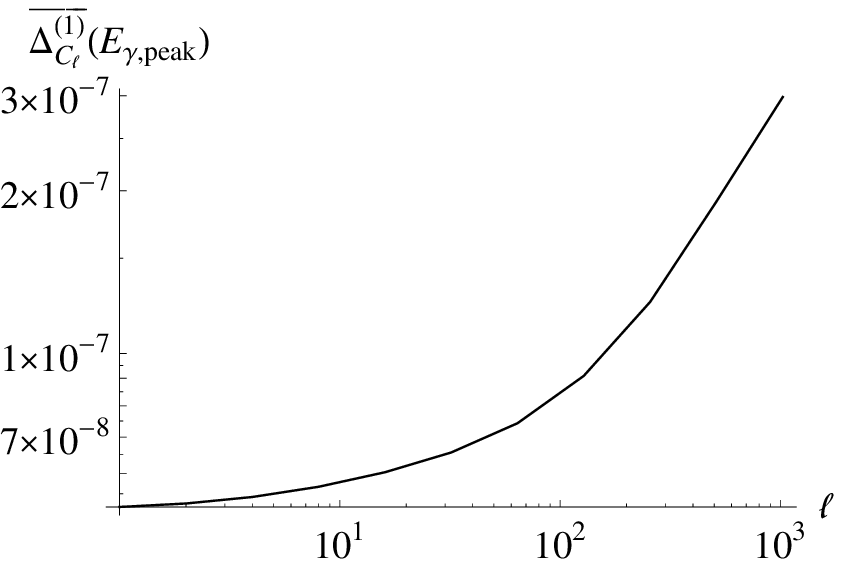}}\quad
  \subfloat{\includegraphics[width=0.4\textwidth]
    {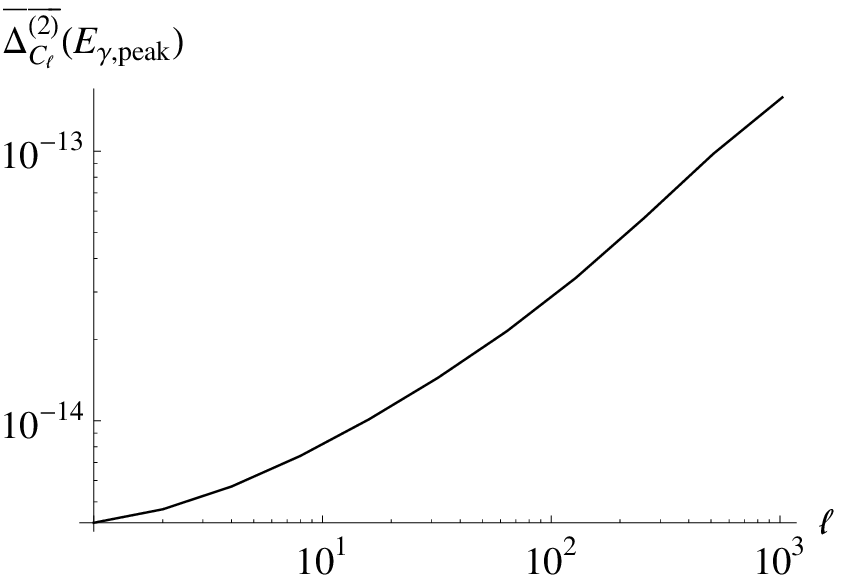}}
  \caption{\label{fig:deltaclbar}\fulljust The coefficients that describe the relative effect of p-wave on the angular power spectrum, according to Eq.~(\ref{eq:deltaclbar}). \hfill\mbox{}}
\end{figure*}

\begin{figure*}
  \subfloat{\includegraphics[width=0.6\textwidth]{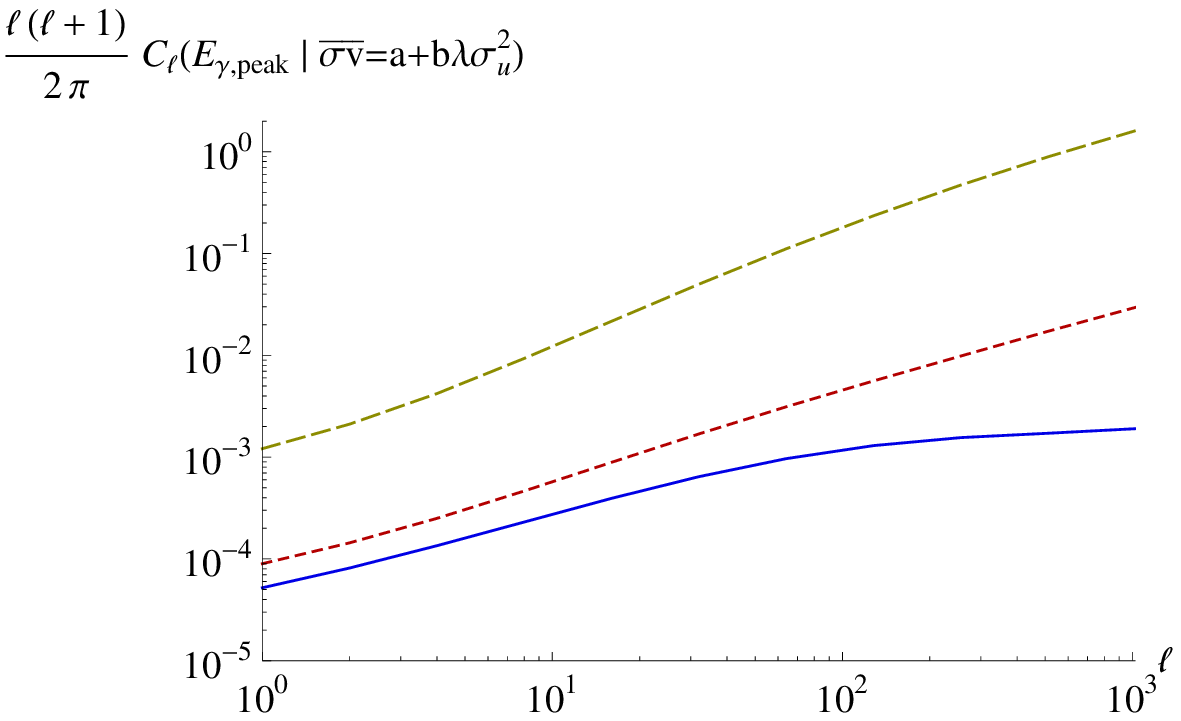}}
  \subfloat{\includegraphics{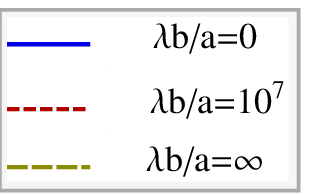}}
  \caption{\label{fig:cl}\fulljust The angular power spectrum of extragalactic, diffuse gamma-rays from dark matter annihilation with different p-wave components. \hfill\mbox{}}
\end{figure*}

To see the effects of the p-wave on the angular power spectrum at the peak energy of the mean intensity spectrum, we plot the $\overline{\Delta_{C_\ell}}$ coefficients in Figure~\ref{fig:deltaclbar}. Here, we have chosen the same model as we mentioned above. If we had chosen the stau neutralino coannihlation scenario, we get a similar picture. These coefficients were nearly universal for the various particle physics models we explored, varying at most by about $50\%$ from the model shown here. Based on the magnitude of the coefficients, we again find that a p-wave will only have a significant effect on the angular power spectrum if $\lambda b/a\agt10^7$; that is, if $b/a\agt10^6$. Unfortunately, the p-wave suppression of these thermal relic theories is so large that it makes it unlikely to observe such a model via indirect detection of extragalactic gamma-rays \cite{Campbell:2010xc}. It may, however, be possible to consider non-thermal relics with both significant s-wave components and strong p-wave strengths. It is interesting to take the general shapes of $\overline{\Delta^{(1)}_{C_\ell}}$ and $\overline{\Delta^{(2)}_{C_\ell}}$, and put them into Eq.~(\ref{eq:deltaclbar}) for various values of $\lambda b/a$ to see how the angular power spectrum can be affected by the coupling of dark matter annihilation to the particle velocity distribution. The results of this exercise are shown in Figure~\ref{fig:cl}. At $b=0$, the usual s-wave angular power spectrum seen in previous works is reproduced \cite{Ando:2005xg,*Ando:2006cr,*Cuoco:2007sh,*SiegalGaskins:2008ge}. We note that a strong p-wave can significantly increase power, more so for large values of $\ell$. If a component of gamma-rays of extragalactic origin is determined to have an angular power spectrum that is best described by a dark matter annihilation with significant $v^2$ component in its cross section, it would be an interesting challenge to understand the mechanisms that allow such a signal to be observable. The magnitude of the effects for the p-wave cross section are very motivating for the consideration of other interesting scenarios, such as annihilation resonances at low particle velocities.

\section{Conclusions}
\label{sec:conclusions}

Using the formalism of the universal spherical halo model of cosmic dark matter, we have developed techniques for estimating the mean intensity and angular power spectrum of gamma-rays due to extragalactic annihilating dark matter for general velocity-dependent annihilation cross section. The formalism for these calculations has been simplified in the case of the mean intensity of the signal, and is new for the angular power spectrum. It is found that the important spatially-dependent field that determines the distribution of intensity is the quantity $\rho^2\overline{\sigma v}$, where $\rho$ is the dark matter density, and $\overline{\sigma v}$ is the locally averaged, relative-velocity-weighted, dark matter annihilation cross section. While the mean intensity can be numerically calculated in a straight-forward way, the angular power spectrum is numerically challenging. It requires evaluation of the Fourier transform of the universal halo profile of $\rho^2\overline{\sigma v}$, which is dependent on the halo's mass $M$ and redshift $z$. Efficient evaluation of the Fourier transform is necessary in order to complete numerical integrations over $M$ and $z$. This is prohibitively time-consuming if a general-purpose integrator is used. We succeeded in developing a sufficient algorithm for this task for the case of rigid NFW halo profiles and annihilation cross sections with a p-wave component.

Thus, we apply our formalism to the specific case of dark matter annihilation with a p-wave component. We find the p-wave only affected the angular power spectrum of extragalactic annihilation gamma-rays for big p-wave strengths, $b/a$, of about $10^6$ or higher, for our model of large scale structure. Since p-wave suppression of intensities of gamma-rays from annihilating thermal relics makes them very difficult to observe, the observation of a component of gamma-rays with an angular power spectrum consistent with strong p-wave annihilations would be an indication of interesting new early-universe physics, and/or new enhancement mechanisms of the annihilation intensity.

Having established that velocity-coupled annihilation of cosmic dark matter can have significant effects on the angular power spectrum of produced gamma-rays, now it will be interesting to understand the angular power spectra associated with other realistic features of annihilation cross-sections, especially those which would enhance the annihilation intensity. Breit-Wigner resonances near the dark matter particle rest mass would produce interesting velocity-dependent effects in the present distribution of cosmic dark matter. Also, Sommerfeld enhancement of annihilation, and Sommerfeld resonances result in new shapes of $\sigma v$ that favor slow moving particles, as opposed to the p-wave, which favors the annihilation in regions where particle are moving rapidly with respect to one another. Our general formalism can be applied to all these cases.

It is likely that halo substructure has a significant effect on the annihilation signal, perhaps increasing the extragalactic mean intensity by a factor as high as 100, as opposed to the intensity from annihilations within our galactic halo being increased by a factor of a few. If, for different halos, this intensity factor varies little with halo mass, it would not affect the angular power spectrum, but significant halo mass dependence would have interesting effects. It is important to add realistic descriptions of the substructure to this formalism to improve its predictive power. Therefore, this will be considered in future work. It is also important for further research to understand the robustness of these calculations against the uncertainties of the large scale structure, in order to have the ability to extract particle information from an observable signal, but as well, to understand how effectively a signal will constrain the properties of the large scale structure distribution.

\begin{acknowledgments}
We thank Eiichiro Komatsu for useful discussions throughout this project. This work is supported in part by DOE Grant DE-FG02-95ER40917. 
\end{acknowledgments}

\appendix

\section{The intensity of annihilation gamma-rays}
\label{ap:intensity}
In a gas of particles (with number density $n$) that may annihilate with one another, the cross section of annihilation $\sigma$ is defined as the rate $\Gamma_p$ of annihilations per target particle divided by the incident flux on the target $nv$ where $v$ is the relative velocity of the incident particle and target particle.  The mean annihilation rate per target at a given position is $\overline{\Gamma}_p=n\overline{\sigma v}$. The annihilation rate per unit volume at a given position is $\Gamma=\frac{1}{2}n\overline{\Gamma}_p=\frac{1}{2}n^2\overline{\sigma v}$.

The rate of energy emitted due to annihilations in a given density of dark matter is given by the power emissivity: the emmitted energy of photons with energy between $E_\gamma$ and $E_\gamma + \der E_\gamma$ per unit volume, time, and energy range $\der E\gamma$.
\[ P_\gamma(E_\gamma)\der E_\gamma=\frac{1}{2}n^2\overline{\sigma v}E_\gamma \der N_\gamma \]
where $\der N_\gamma$ is the number of number of photons per annihilation with energy between $E_\gamma$ and $E_\gamma+\der E_\gamma$. We call $\frac{\der N_\gamma}{\der E_\gamma}$ the differential photon spectrum per annihilation.

In a flat Friedmann-Robertson-Walker cosmology, using coordinates in the cosmological rest frame, the proper volume of space with solid angle $\der\Omega$ and thickness $\der z$ at redshift $z$ is 
\[ \der V=[a(z)\der r][a^2(z)r^2\der\Omega]=\frac{1}{(1+z)^3}r^2\der r\der\Omega\]
where $a$ is the cosmological scale parameter.

The luminosity $\der L_\gamma(E_\gamma)$ of photons of energy $E_\gamma$ emitted from this region of space because of annihilations is
\[ \der L_\gamma(E_\gamma,z)=P_\gamma(E_\gamma,z)\der V=\frac{1}{2} n^2 \overline{\sigma v}E_\gamma\frac{\der N_\gamma}{\der E_\gamma}(E_\gamma) \frac{1}{(1+z)^3}r^2\der r\der\Omega.\] Assuming isotropic emission, the photons emitted by this volume pass with uniform flux density through any sphere centered on the source. The sphere on which we sit, centered on the source, has proper surface area \[A=4\pi r^2a^2(0)=4\pi r^2.\] The total luminosity on this shell (energy of photons emitted from the source with energies between $E_\gamma$ and $E_\gamma+\der E_\gamma$, per $\der E_\gamma$, per unit time of emission) is redshifted: the cosmological redshift of photon energy due to the expansion of the universe is cancelled by the redshift of the energy bin $\der E_\gamma$; the arrival rate of photons is redshifted giving one factor of $(1+z)^{-1}$. Observation of photons of energy $E_\gamma$ means photons of energy $(1+z)E_\gamma$ were emitted. Hence, the luminosity of photons on the observer's spherical shell with energy $E_\gamma$ from the source at redshift $z$ is
\[\der L'_\gamma(E_\gamma,z)=\frac{\der L_\gamma((1+z)E_\gamma,z)} {1+z} e^{-\tau((1+z)E_\gamma,z)}\]
where $\tau(E_\gamma,z)$ is the opacity of the universe to gamma rays \cite{Stecker:2005qs,*Stecker:2006eh}. The photon flux on the sphere, or surface brightness, due to a source at position $\mathbf{r}$ and redshift $z$ is
\[\der S_\gamma(E_\gamma,\mathbf{r},z)=\frac{\der L'_\gamma(E_\gamma,\mathbf{r},z)} {A(z)} = \frac{1}{8\pi}\frac{1}{(1+z)^4} n^2(\mathbf{r})\overline{\sigma v}(\mathbf{r}) (1+z)E_\gamma \frac{\der N_\gamma} {\der E_\gamma}((1+z)E_\gamma) e^{-\tau((1+z)E_\gamma,z)}\der r\der\Omega.\] The net specific intensity (number of photons of energy $E_\gamma$ observed per bin $\der E_\gamma$, per unit time, per source solid angle, per normal photon collecting area) is found from a line-of-sight integration in direction $\hat{\mathbf{n}}$:
\begin{equation}
  \label{eq:inten}
  I_\gamma(E_\gamma,\hat{\mathbf{n}})=\int\frac{\der S_\gamma(E_\gamma,z)}
  {E_\gamma\der\Omega}=\int\frac{\der z}{H(z)}W((1+z)E_\gamma,z)
  [\rho^2 \overline{\sigma v}](\hat{\mathbf{n}},z)
\end{equation}
where the coordinate distance is related to redshift via $\der r=-\frac{\der z} {H(z)}$ with the usual Hubble function given by $H(z)=H_0\sqrt{\Omega_m(1+z)^3+\Omega_\Lambda}$ in terms of the Hubble constant $H_0$, local matter abundance $\Omega_m$, and local dark energy abundance $\Omega_\Lambda=1-\Omega_m$. The important spatially dependent field $\rho^2\overline{\sigma v}$, where $\rho$ is the density of dark matter, is weighted by the intensity window function 
\begin{equation}
  W(E_\gamma,z)=\frac{1}{8\pi m^2_{\text{DM}}}\frac{1}{(1+z)^3}\frac{\der
    N_\gamma}{\der E_\gamma}(E_\gamma)e^{-\tau(E_\gamma,z)}
\end{equation}
with $m_{\text{DM}}$ being the mass of the dark matter particle.

\section{Modeling large scale structure with the spherical halo model}
\label{ap:lss}
Statistical properties of the spatial distribution of $\rho$ and $\overline{\sigma v}$ at each redshift can be modeled with the disjoint spherical halo model. Since $\rho$ throughout the universe is greatest at the cores of halos that are nearly spherical, it is reasonable to assume that the dominant contribution to an annihilation signal is due to dark matter consisting of an ensemble of disjoint spherical halos, each at position $\mathbf{R}_i$ with some mass $M_i$. 
\subsection{The point distribution of halos}
The point distribution of $N_h$ halos at redshift $z$ is
\[p_h(\mathbf{r},M,z)=\sum_{i=1}^{N_h(z)}\delta^{(3)}(\mathbf{r}-\mathbf{R}_i(z)) \delta(M-M_i(z)).\]
The halo mass function is the ensemble average of the point distribution:
\[\frac{\der n}{\der M}(M,z)=\mean{p_h}\!(M,z)\]
normalized such that
\[\int\der M\frac{\der n}{\der M}M=\mean{\rho}\!(z)=\rho_c\Omega_{\text{DM}}(1+z)^3\]
where $\rho_c=\frac{3H_0^2}{8\pi G}$ is the local cosmological critical density, and $\Omega_{\text{DM}}$ is the local cold dark matter content.

The halo overdensity can be defined formally as 
\[\delta_h(\mathbf{r},M,z)=\frac{p_h(\mathbf{r},M,z)}{\frac{\der n}{\der M} (M,z)}-1\]
which has zero ensemble average. This is correlated to the matter overdensity
\[\delta_\rho(\mathbf{r},z)=\frac{\rho(\mathbf{r},z)}{\mean{\rho}\!(z)}-1.\]
This correlation is loosely described by the halo bias function
\[ b(M,z)\sim\mean{\frac{\delta_h(\mathbf{r},M,z)}{\delta_\rho(\mathbf{r},z)}}, \]
which will be described, for our purposes, more precisely below in terms of the halo power spectrum.

The matter correlation function over 2 positions $\mathbf{r}_1$ and $\mathbf{r}_2$ at the same redshift $z$ is $\xi(\mathbf{r}_1,\mathbf{r}_2,z)= \mean{\delta_\rho(\mathbf{r}_1,z)\delta_\rho(\mathbf{r}_2,z)}$. It only depends on the separation $r=|\mathbf{r}_1-\mathbf{r}_2|$ of the 2 points: $\xi(r,z)= \mean{\delta_\rho(\mathbf{r}_1,z)\delta_\rho(\mathbf{r}_1+\mathbf{r},z)}$. The power spectrum is the Fourier transform of the correlation function: $P_{\rho,\rho}(k,z)=\int\der^3\mathbf{r}e^{-i\mathbf{k}\cdot\mathbf{r}}\xi(r,z)$. The evolution equations for $\delta_\rho$ are non-linear. When linearized, the linear solution results in a power spectrum valid on linear scales called the linear power spectrum.
\[ P_{\rho,\rho}(k,z)\approx P_{\text{lin}}(k,z) \qquad\text{for small $k$.}\]

In analogy, the full halo correlation function is $\tilde{\xi}_h(\mathbf{r}_1, M_1, \mathbf{r}_2, M_2, z)=\mean{\delta_h(\mathbf{r}_1,M_1,z) \delta_h(\mathbf{r}_2,M_2,z)}.$ We can see that this function has a singularity at $\mathbf{r}_1=\mathbf{r}_2$ by expanding
\[ \tilde{\xi}_h(\mathbf{r}_1, M_1, \mathbf{r}_2, M_2, z) = \frac{\mean{p_h(\mathbf{r}_1, M_1, z)p_h(\mathbf{r}_2, M_2, z)}}{\frac{\der n}{\der M}(M_1,z)\frac{\der n}{\der M}(M_2,z)}-1\]
and splitting the point distribution 2-moment into diagonal and non-diagonal pieces.
\begin{eqnarray}
  \nonumber
  \langle p_h(\mathbf{r}_1, M_1, z)&p_h&(\mathbf{r}_2, M_2, z)\rangle
    =\mean{\sum_{i=1}^{N_h}\sum_{j=1}^{N_h}
    \delta^{(3)}(\mathbf{r}_1-\mathbf{R}_i)\delta(M_1-M_i)
    \delta^{(3)}(\mathbf{r}_2-\mathbf{R}_j)\delta(M_2-M_j)}\\
  \nonumber
  &=&\mean{\sum_i\sum_{j\neq i}
    \delta^{(3)}(\mathbf{r}_1-\mathbf{R}_i)\delta(M_1-M_i)
    \delta^{(3)}(\mathbf{r}_2-\mathbf{R}_j)\delta(M_2-M_j)}\\
  \nonumber
  &&+ \mean{\sum_i\delta^{(3)}(\mathbf{r}_1-\mathbf{R}_i)\delta(M_1-M_i)
    \delta^{(3)}(\mathbf{r}_2-\mathbf{R}_i)\delta(M_2-M_i)}\\
  \nonumber
  &=&C_h^{(2)}(\mathbf{r}_1, M_1, \mathbf{r}_2, M_2, z) + 
    \delta^{(3)}(\mathbf{r}_1-\mathbf{r}_2)\delta(M_1-M_2)
    \frac{\der n}{\der M}(M_2,z)
\end{eqnarray}
This expression makes the singularity explicit. Here, the function $C_h^{(2)}$ is introduced as the non-diagonal part of the halo 2-moment. Note that if one restricts to disjoint halo ensembles, then $C_h^{(2)}=0$ formally.

The non-diagonal part of the halo correlation function, known simply as the halo correlation function \footnote{When describing $\tilde{\xi}_h$, we refer to it as the \emph{full} halo correlation function to differentiate it from $\xi_h$.}, is now everywhere finite and is defined as
\[ \xi_h(\mathbf{r}_1, M_1, \mathbf{r}_2, M_2, z) \equiv
  \frac{C_h^{(2)}(\mathbf{r}_1,M_1,\mathbf{r}_2,M_2,z)}
  {\frac{\der n}{\der M}(M_1,z)\frac{\der n}{\der M}(M_2,z)}-1.\]
Then
\[ \tilde{\xi}_h(\mathbf{r}_1, M_1, \mathbf{r}_2, M_2, z) = 
  \xi_h(\mathbf{r}_1, M_1, \mathbf{r}_2, M_2, z) +
  \frac{\delta^{(3)}(\mathbf{r}_1-\mathbf{r}_2)\delta(M_1-M_2)}
  {\frac{\der n}{\der M}(M_1,z)}.\]

The halo power spectrum is defined accordingly as
\[ P_{h,h}(k, M_1, M_2, z)\equiv\int\der^3\mathbf{r} 
  e^{-i\mathbf{k}\cdot\mathbf{r}}
  \xi_h(\mathbf{r}_1, M_1, \mathbf{r}_1+\mathbf{r}, M_2, z).\]
Its correlation to the matter power spectrum is encoded in the halo bias function. For our purposes, the precise meaning of the halo bias is through the relation
\[ P_{h,h}(k, M_1, M_2, z)=b(M_1,z)b(M_2,z)P_{\rho,\rho}(k,z). \]
Although recent works show it is likely that these bias functions have some dependence on scale $k$ \cite{Tseliakhovich:2010bj,*Shandera:2010ei,*Parfrey:2010uy}, we did not consider such models in this work. It turns out that the halo power spectrum is only relevant for us on linear scales where $P_{\rho,\rho}$ may be substituted with the linear power spectrum. For less abstract and more physical discussion of many of the quantities in this section, we recommend Peebles (1980) \cite{Peebles:1980} where the discussion is in terms of galaxies instead of dark matter halos, but is still relevant none-the-less.

For our calculations, the knowledge we require of the halo distribution is the mass function and halo bias function. For this work, we used the Sheth-Tormen halo mass function \cite{Sheth:2001dp} and the halo bias due to Sheth, Mo, and Tormen \cite{Sheth:1999su}. For the matter distribution, we used the best-fit cosmological parameters from WMAP5 \cite{Komatsu:2008hk} and the linear power spectrum of Eisenstein and Hu \cite{Eisenstein:1997jh} with neutrino free-streaming and gravitational wave effects neglected. Dark matter halos were taken to have a minimum mass cutoff of $10^{-6} M_\odot$.

\subsection{Universal Halo Profiles}

In this section, we describe radial profiles $F_h(r|M,z)$ of halos that are seen to be universal in large N-body simulations of self-gravitating, collisionless particles. These radial profiles in our model are written to depend only on the halo's mass $M$ and observed redshift $z$. It is possible that, in the future, certain variations between halos of same mass and redshift might be taken into account by extending the universality to depend on other halo properties (such as halo concentration, formation redshift, angular momentum, shape parameters, etc.), and appropriately extending the halo mass and bias functions to also depend on those halo variables.

The first hint that the phase space of relaxed dark matter halos is universally stratified came when Navarro, Frenk, and White \cite{Navarro:1996gj} observed that the spherically-averaged radial density profiles of the halos in their simulations were well-described by what is now known as the NFW profile
\begin{equation}
  \label{eq:nfw}
  \rho_h(r|\rho_s,r_s,R_{\text{vir}})=
  \begin{cases}
    \frac{\rho_s}{\frac{r}{r_s}\left(1+\frac{r}{r_s}\right)^2}& \text{for $r<R_{\text{vir}}$},\\
    0& \text{otherwise.}
  \end{cases}
\end{equation}
Halo models typically truncate the halo profile artificially at a virial radius $R_{\text{vir}}(M,z)$ of the mass $M$ halo located at redshift $z$, which is fine for our calculations. Various mass-radius relations have been used in the literature and the effect of many of them on the halo mass function were studied in \cite{White:2002at}. Based on the results of that study, we adopted the convention
\[ M=\frac{4}{3}\pi R^3_{\text{vir}}(M,z)\,\Delta_{\text{vir}}\mean{\rho}\!(z)\]
with $\Delta_{\text{vir}}=180$. Although there are more modern density profiles that better describe the most recent simulations \cite{Navarro:2008kc}, the NFW profile still provides a reasonable description of simulated halos and is a good place to begin development of techniques for numerical calculation because of its simple analytic form. The usual method for writing (\ref{eq:nfw}) in its universal form is to use our mass-radius relation to define a halo concentration
\[ c(M,z)\equiv\frac{R_{\text{vir}}(M,z)}{r_s(M,z)} \]
which we take to be distributed according to the model in \cite{Bullock:1999he}. This determines $r_s(M,z)$. Solving for $M$ by integrating over the halo density requires
\[ \rho_s(c,z)=\frac{\Delta_{\text{vir}}\overline{\rho}(z)c^3} 
  {3\left[\ln(1+c)-\frac{c}{1+c}\right]} \]
which fixes $\rho_s(M,z)$. Thus, we determine the NFW density profile distribution $\rho_h(r|M,z)$.

It is supposed that the distribution of particle velocities $\mathbf{u}$ at each position $r$ of a relaxed universal halo is well described by a velocity distribution function $f_{h,\mathbf{u}}(\mathbf{u}|r,M,z)$. The halo's velocity dispersion $\sigma_{uh}$, observed in simulations to be nearly isotropic at the halo cores and taken as such in this work, is fixed by the observation of stratification of the so-called pseudo-phase-space-density in simulations according to \cite{Taylor:2001bq}
\[ \frac{\rho_h}{\sigma_{uh}^3}\propto r^{-\alpha}. \]
Coupled with the radial Jeans equation, which describes the phase space evolution of self-gravitating collisionless systems, the Dehnen-McLaughlin (D-M) halo profiles and velocity dispersions are determined \cite{Dehnen:2005cu}. Their critical, NFW-like solution has $\alpha=\frac{35}{18}$, consistent with values of the radial dispersion seen in the simulations \cite{Navarro:2008kc}. The scale radius $r_0$ for these profiles is at the position where $-\der\ln\rho_h(r_0)/\der\ln r=6-2\alpha$. Matching the NFW profile to the critical D-M profile at $r_0$, we find an NFW velocity variance halo profile given by
\[ \sigma_{uh}^2(r)=\frac{\sigma_s^2\left(\frac{r}{r_s}\right)^\beta}
  {\left(1+\frac{r}{r_s}\right)^{4/3}} \]
where $\beta=\frac{2}{3}(\alpha-1)=\frac{17}{27}$, the scale variance is
\[ \sigma_s^2 = 12^{2/3}\pi G\kappa^{-1}
  \left(\beta-\frac{1}{3}\right)^{\beta-1/3}\left(1-\beta\right)^{1-\beta}
  \rho_s r_s^2, \]
and $\kappa=\frac{200}{81}$ for the critical D-M solution.

What we are specifically interested in for the annihilation cross section is the distribution of particle relative velocities $v$ at a given position. If one knows $f_{h,\mathbf{u}}(\mathbf{u}|r,M,z)$ then the 2-particle distribution of, say, square relative velocities $f_{h,v^2}(v^2|r,M,z)$ is easily determined. In the case where $f_{h,\mathbf{u}}$ is Maxwell-Boltzmann, one finds the mean square relative velocity is simply
\[ \overline{v^2}=6\sigma_{u}^2, \]
a relation that would hold at each position of the halo. Since the velocity distribution is well-established to deviate from Maxwell-Boltzmann \cite{Kazantzidis:2003im,*Wojtak:2005fe,*Hansen:2005yj,*Kuhlen:2009vh}, one would expect a generalized relation such as
\[ \overline{v_h^2}(r|M,z)=\lambda(r|M,z)\sigma_{uh}^2(r|M,z) \]
to hold in relaxed spherical halos. For this work, we will continue by approximating $\lambda$ to be constant and treat it as a free parameter of our model, to eventually be determined from simulation data. We used $\lambda=6$ for calculations that required a specific value.

Taking the velocity-weighted dark matter annihilation cross section $[\sigma v](v^2)$ as a function of $v^2$, the halo's mean cross section profile is determined
\[ [\overline{\sigma v}]_h(r|M,z)=\int \der[v^2]\,[\sigma v](v^2) 
  f_{h,v^2}(v^2|r,M,z). \]
For p-wave annihilation with $\sigma v=a+bv^2$, this process is trivial.
\begin{equation}
  \label{eq:pwavehalo}
  [\overline{\sigma v}]_h(r|M,z)=a+b\lambda\sigma_{uh}^2(r|M,z)
\end{equation}

\section{The mean intensity and angular power spectrum of extragalactic gamma-rays: application of large scale structure}
\label{ap:signals}

From (\ref{eq:inten}), the mean intensity of annihilation gamma-rays is found simply from averaging over ensembles of dark matter halos
\begin{equation}
  \mean{I_\gamma}(E_\gamma)=\int\frac{\der z}{H(z)}
  W((1+z)E_\gamma,z)\mean{\rho^2\overline{\sigma v}}\!(z).
\end{equation}
In the disjoint halo model, an ensemble of halos at redshift $z$ has
\[ [\rho^2\overline{\sigma v}](\mathbf{r},z) = 
  \sum_{i=1}^{N_h(z)}\rho_h^2(\mathbf{r}-\mathbf{R}_i(z)\,|\,M_i(z),z)\ 
  [\overline{\sigma v}]_h\!(\mathbf{r}-\mathbf{R}_i(z)\,|\,M_i(z),z) \]
where $\mathbf{r}$ are a global set of coordinates at the time associated with redshift $z$. For disjoint ensembles, at most one term contributes to the sum at any given position $\mathbf{r}$. This allows us to express the ensemble average in terms of the halo mass function.
\begin{eqnarray}
  \nonumber
  \mean{\rho^2\overline{\sigma v}}\!(z)&=&\mean{\int\der^3\mathbf{R}\,\der M\,
    \rho_h^2(\mathbf{r}-\mathbf{R}\,|\,M,z)\ 
    [\overline{\sigma v}]_h\!(\mathbf{r}-\mathbf{R}\,|\,M,z)
    \sum_{i=1}^{N_h(z)}\delta^{(3)}(\mathbf{R}-\mathbf{R}_i(z))\,
    \delta(M-M_i(z))} \\
  &=&\int\der^3\mathbf{R}\,\der M\frac{\der n}{\der M}(M,z)\ 
    \rho_h^2(R|M,z)\ [\overline{\sigma v}]_h\!(R|M,z)
\end{eqnarray}

We explore the angular anisotropies in the intensity signal by determining its angular power spectrum defined as
\[ C_\ell=\mean{|a_{\ell m}|^2} \]
with spherical harmonic coefficients obtained from
\[ \delta_I(\hat{\mathbf{n}},E_\gamma)\equiv\frac{I_\gamma(\hat{\mathbf{n}},
  E_\gamma)}{\mean{I_\gamma}(E_\gamma)}-1=\sum_{\ell=0}^\infty
  \sum_{m=-\ell}^\ell a_{\ell m}(E_\gamma)Y_{\ell m}(\hat{\mathbf{n}}),\]
or
\begin{eqnarray}
  \nonumber
  a_{\ell m}(E_\gamma)&=&\oint\der\Omega\,\delta_I(\hat{\mathbf{n}},E_\gamma)\,
    Y^*_{\ell m}(\hat{\mathbf{n}}).\\
  \nonumber
  &=&\frac{1}{\mean{I_\gamma}\!(E_\gamma)}\oint\der\Omega\,\int\frac{\der z}
    {H(z)}\big\{\,[\rho^2\overline{\sigma v}](\hat{\mathbf{n}},z)-
    \mean{\rho^2\overline{\sigma v}}(z)\big\}\,W((1+z)E_\gamma,z)\,
    Y^*_{\ell m}(\hat{\mathbf{n}})\\
  \nonumber
  &=&\frac{1}{\mean{I_\gamma}\!(E_\gamma)}\int\frac{\der z}{H(z)}
    \mean{\rho^2\overline{\sigma v}}\!(z)\,W((1+z)E_\gamma,z)
    \int\der\Omega\,\delta_{\rho^2\overline{\sigma v}}(\hat{\mathbf{n}},z)
    Y^*_{\ell m}(\hat{\mathbf{n}})
\end{eqnarray}
where, as usual,
\[ \delta_{\rho^2\overline{\sigma v}}\equiv\frac{\rho^2\overline{\sigma v}}
  {\mean{\rho^2\overline{\sigma v}}}-1.\]
Then
\[ C_\ell(E_\gamma)=\frac{1}{\mean{I_\gamma}^2\!(E_\gamma)}\int\frac{\der z}
  {H(z)}\frac{\der z'}{H(z')}\mean{\rho^2\overline{\sigma v}}\!(z)
  \mean{\rho^2\overline{\sigma v}}\!(z')\,W\!((1+z)E_\gamma,z)\,
  W\!((1+z')E_\gamma,z')\,F_\ell(z,z') \]
where
\[ F_\ell(z,z')\equiv\int\der\Omega\der\Omega'\mean{\delta_{\rho^2
  \overline{\sigma v}}(\hat{\mathbf{n}},z)\delta_{\rho^2\overline{\sigma v}}
  (\hat{\mathbf{n}}',z')}Y^*_{\ell m}(\hat{\mathbf{n}})
  Y_{\ell m}(\hat{\mathbf{n}}'). \]
We'll see shortly why $F_\ell$ is independent of $m$. To simplify it, write in terms of the power spectrum of the $\rho^2\overline{\sigma v}$ field,
\[ \mean{\delta_{\rho^2\overline{\sigma v}}(\hat{\mathbf{n}},z)
  \delta_{\rho^2\overline{\sigma v}}(\hat{\mathbf{n}}',z')}=
  \int\frac{\der^3 \mathbf{k}}{(2\pi)^3}e^{irk\hat{\mathbf{n}}\cdot\hat{\mathbf{k}}}
  e^{-ir'k\hat{\mathbf{n}'}\cdot\hat{\mathbf{k}}}P_{\rho^2\overline{\sigma v}}(k,z,z'), \]
where 
\[ r=\int_0^z\frac{\der\overline{z}}{H(\overline{z})} \]
is the distance to redshift $z$, and similarly for $r'$. Applying Rayleigh's formula
\[ e^{irk\hat{\mathbf{n}}\cdot\hat{\mathbf{k}}}=4\pi\sum_{\ell'=0}^{\infty}
  \sum_{m'=-\ell'}^{\ell'}i^{\ell'}j_{\ell'}(kr)Y^*_{\ell' m'}(\hat{\mathbf{k}})
  Y_{\ell' m'}(\hat{\mathbf{n}}) \]
and the orthogonality of spherical harmonics, one finds
\[ F_\ell(z,z')=\frac{2}{\pi}\int_0^\infty\der k k^2
  P_{\rho^2\overline{\sigma v}}(k,z,z')j_\ell(kr)j_\ell(kr'). \]
One would not expect any significant correlation between regions of different redshift along a line-of-sight. One way this is realized is when $P_{\rho^2\overline{\sigma v}}$ is a slowly-varying function of $k$. In this case, it is a good approximation to treat it as a constant at wave number where $j_\ell(kr)$ is maximized. Since $j_\ell(x)$ has its maximum near $x=\ell$, we can approximate the power spectrum by its value at $k=\ell/r(z)$. Then orthogonality of the spherical Bessel functions
\[ \int_0^\infty\der k k^2j_\ell(kr)j_\ell(kr')=\frac{\pi}{2r^2}\delta(r-r')=
  \frac{\pi}{2r^2(z)}H(z)\delta(z-z') \]
gives
\[ F_\ell(z,z')\approx
  \delta(z-z')\frac{H(z)}{\ell^2}\,k^2P_{\rho^2\overline{\sigma v}}(k,z)
  \Big |_{k=\frac{\ell}{r(z)}}. \]
We then finally express the angular power spectrum as
\begin{equation}
  C_\ell(E_\gamma)\approx\frac{1}{\ell^2\mean{I_\gamma}^2\!\!(E_\gamma)}
    \int\frac{\der z}{H(z)}W^2((1+z)E_\gamma,z)k^2
    \overline{P}_{\rho^2\overline{\sigma v}}(k,z)\Big |_{k=\frac{\ell}{r(z)}},
\end{equation}
where we denote
\[ \overline{P}_{\rho^2\overline{\sigma v}}(k,z)\equiv\mean{\rho^2
  \overline{\sigma v}}^2\!(z)P_{\rho^2\overline{\sigma v}}(k,z). \]
To derive the expression for the power spectrum of $\rho^2\overline{\sigma v}$, consider the correlation function at two points $\mathbf{r}_1,\mathbf{r}_2$ at the same redshift $z$.
\[ \mean{\delta_{\rho^2\overline{\sigma v}}(\mathbf{r}_1,z)\delta_{\rho^2\overline{\sigma v}}(\mathbf{r}_2,z)}=
  \frac{\mean{[\rho^2\overline{\sigma v}](\mathbf{r}_1,z)[\rho^2\overline{\sigma v}](\mathbf{r}_2,z)}}
  {\mean{\rho^2\overline{\sigma v}}^2\!(z)}-1 \]
Recalling
\[ \mean{p_h(\mathbf{R}_1,M_1,z)p_h(\mathbf{R}_2,M_2,z)}=\frac{\der n}{\der M}(M_1,z)
  \frac{\der n}{\der M}(M_2,z)[\tilde{\xi}_h(\mathbf{R}_1,M_1,\mathbf{R}_2,M_2,z)+1], \]
the 2-moment becomes
\begin{eqnarray}
  \nonumber
  \Big\langle&&\!\!\!\!\!\!\!\![\rho^2\overline{\sigma v}](\mathbf{r}_1,z)[\rho^2\overline{\sigma v}]
    (\mathbf{r}_2,z)\Big\rangle=\mean{\sum_i\sum_j[\rho^2\overline{\sigma v}]_h
    (\mathbf{r}_1-\mathbf{R}_i|M_i,z)[\rho^2\overline{\sigma v}]_h
    (\mathbf{r}_2-\mathbf{R}_j|M_j,z)}\\
  \nonumber
  &=&\int\der^3\mathbf{R}_1\der M_1\der^3\mathbf{R}_2\der M_2
  	[\rho^2\overline{\sigma v}]_h(\mathbf{r}_1-\mathbf{R}_1|M_1,z)
  	[\rho^2\overline{\sigma v}]_h(\mathbf{r}_2-\mathbf{R}_2|M_2,z)
    \mean{p_h(\mathbf{R}_1,M_1,z)p_h(\mathbf{R}_2,M_2,z)}\\
  \nonumber
  &=&\int\der^3\mathbf{R}_1\der M_1\der^3\mathbf{R}_2\der M_2
    \frac{\der n}{\der M}(M_1,z)\frac{\der n}{\der M}(M_2,z)
    [\rho^2\overline{\sigma v}]_h(\mathbf{r}_1-\mathbf{R}_1|M_1,z)
    [\rho^2\overline{\sigma v}]_h(\mathbf{r}_2-\mathbf{R}_2|M_2,z)\\
  \nonumber
  &&\times\ \xi_h(\mathbf{R}_1,M_1,\mathbf{R}_2,M_2,z)
    \ +\int\der^3\mathbf{R}\der M\frac{\der n}{\der M}(M,z)
    [\rho^2\overline{\sigma v}]_h(\mathbf{r}_1-\mathbf{R}|M,z)
    [\rho^2\overline{\sigma v}]_h(\mathbf{r}_2-\mathbf{R}|M,z)
    +\mean{\rho^2\overline{\sigma v}}^2\!(z)
\end{eqnarray}
where the second term in the last line is due to the singularity in $\tilde{\xi}_h$. We therefore find the correlation function to be
\begin{eqnarray}
  \nonumber
  &&\mean{\delta_{\rho^2\overline{\sigma v}}(\mathbf{r}_1,z)\delta_{\rho^2\overline{\sigma v}}(\mathbf{r}_2,z)}=
    \int\der^3\mathbf{R}\der M\frac{\der n}{\der M}(M,z)\frac{[\rho^2\overline{\sigma v}]_h(\mathbf{r}_1-\mathbf{R}|M,z)
    [\rho^2\overline{\sigma v}]_h(\mathbf{r}_2-\mathbf{R}|M,z)}{\mean{\rho^2\overline{\sigma v}}^2\!(z)}\\
  \nonumber
  &&\quad\quad+\int\der^3\mathbf{R}_1\der M_1\der^3\mathbf{R}_2\der M_2
    \frac{\der n}{\der M}(M_1,z)\frac{\der n}{\der M}(M_2,z)
    \frac{[\rho^2\overline{\sigma v}]_h(\mathbf{r}_1-\mathbf{R}_1|M_1,z)
    [\rho^2\overline{\sigma v}]_h(\mathbf{r}_2-\mathbf{R}_2|M_2,z)}
    {\mean{\rho^2\overline{\sigma v}}^2\!(z)}\\
  \nonumber
  &&\quad\quad\quad\times\ \xi_h(\mathbf{R}_1,M_1,\mathbf{R}_2,M_2,z).
\end{eqnarray}
This simplifies significantly in momentum space. If we determine the Fourier transform of the halo profile
\[ \mathcal{FT}\{[\rho^2\overline{\sigma v}]_h\}(k|M,z)=\int\der^3\mathbf{r}
  e^{-i\mathbf{k}\cdot\mathbf{r}}[\rho^2\overline{\sigma v}]_h(r|M,z), \]
the power spectrum can be written
\begin{eqnarray}
  \nonumber
  \overline{P}_{\rho^2\overline{\sigma v}}(k,z)&=&\mean{\rho^2\overline{\sigma v}}^2\!(z)
    \int\der^3\mathbf{r}\mean{\delta_{\rho^2\overline{\sigma v}}(\mathbf{r}_1,z)
    \delta_{\rho^2\overline{\sigma v}}(\mathbf{r}_1+\mathbf{r},z)}e^{-i\mathbf{r}\cdot\mathbf{k}}\\
  \nonumber
  &=&\int\der M\frac{\der n}{\der M}(M,z)\,\mathcal{FT}\!\{[\rho^2\overline{\sigma v}]_h\}^2\!(k|M,z)\\
  \nonumber
  &&+\int\der M_1\der M_2\frac{\der n}{\der M}(M_1,z)\frac{\der n}{\der M}(M_2,z)\,
    \mathcal{FT}\!\{[\rho^2\overline{\sigma v}]_h\}(k|M_1,z)
    \mathcal{FT}\!\{[\rho^2\overline{\sigma v}]_h\}(k|M_2,z)P_h(k,M_1,M_2,z).
\end{eqnarray}
The first term, the one-halo term, dominates at small scales (large $k$) and the second term, the two-halo term, dominates at the large scales, in the linear regime. So in this expression, it is correct to use
\[ P_h(k,M_1,M_2,z)=b(M_1,z)b(M_2,z)P_{\text{lin}}(k,z). \]
Therefore, the power spectrum is simply expressed as
\begin{eqnarray}
  \nonumber
  \overline{P}_{\rho^2\overline{\sigma v}}(k,z)&=&\int\der M
    \frac{\der n}{\der M}(M,z)\big[\mathcal{FT}\!\{[\rho^2
    \overline{\sigma v}]_h\}\!(k|M,z)\big]^2\\
  &&+\left[\int\der M\frac{\der n}{\der M}(M,z)b(M,z)
    \mathcal{FT}\!\{[\rho^2\overline{\sigma v}]_h\}(k|M,z)\right]^2
    P_{\text{lin}}(k,z).
\end{eqnarray}

\section{Numerical evaluation of needed Fourier transforms of rigid NFW profiles}
\label{ap:ft}
Although there exist some very good general integrators for Fourier transforms \cite{Piessens:1983}, their use is not very feasible in this calculation. The transforms appear in the integrand of the halo mass integration, and that result is then integrated over redshift. The number of evaluations required for precise calculation is very large, and takes too long to complete when using a general-purpose integrator. Since these functions are over a 3-dimensional space $(k|M,z)$ that stretches over a large range of scales, it is also not feasible to fill a data table for interpolation.

For the rigid NFW profile, a closed form solution is available for $\mathcal{FT}\{\rho_h^2\}$, which has allowed efficient calculation of s-wave angular power spectra in previous works. No such closed form is available for the non-analytic $\mathcal{FT}\{\rho_h^2\sigma_{uh}^2\}$. Nevertheless, we were successful in developing an efficient numerical algorithm for efficient evaluation of this function, described below. One of the challenges for calculations of angular power spectra of extragalactic dark matter annihilation products is the development of efficient numerical methods to evaluate $\mathcal{FT}\{\rho_h^2[\overline{\sigma v}]_h\}$ for a given model's halo profiles and annihilation cross section. This calculation would have taken weeks to complete using the quadpack general purpose Fourier transform integrator, qawf. With the algorithm described in this section, the results in this paper were able to be evaluated within a few days of run time on a desktop computer.

\subsection{$\mathcal{FT}\{\rho_h^2\}(k|M,z)$}
\label{subap:den2}
This Fourier transform can be expressed as
\begin{eqnarray}
  \nonumber
  \mathcal{FT}\{\rho_h^2\}(k)&=&4\pi\rho_s^2r_s^3\left\{-\frac{2}{3}+
    \frac{4+3c}{6(1+c)^2}\cos(kr_sc)+\frac{11+15c+6c^2-[(1+c)kr_s]^2}
    {6kr_s(1+c)^3}\sin(kr_sc)+\frac{\text{Si}(kr_sc)}{kr_s}\right.\\
  &&\left.-\begin{pmatrix} 
      1-\frac{(kr_s)^2}{6} & \frac{1}{kr_s}-\frac{kr_s}{2}
    \end{pmatrix}
    \begin{pmatrix}
      \cos(kr_s) & \sin(kr_s) \\
      -\sin(kr_s) & \cos(kr_s)
    \end{pmatrix}
    \begin{pmatrix}
      \text{Ci}\big(kr_s(1+c)\big)-\text{Ci}(kr_s) \\
      \text{Si}\big(kr_s(1+c)\big)-\text{Si}(kr_s)
    \end{pmatrix}\right\}
\end{eqnarray}
where Si and Ci are the sine integral and cosine integral, respectively, for which efficient numerical methods for evaluation already exist \cite{Press:2007}. Evaluating the line-of-sight integrand for the angular power spectrum near $z=0$ requires the Fourier transform to be evaluated in the $k\rightarrow\infty$ regime. One finds that for $kr_s\gg1$,
\begin{equation*}
  \mathcal{FT}\{\rho_h^2\}(k)=2\pi r_s^3\rho_s^2\left\{\frac{\pi}{kr_s}
    -\frac{1}{(kr_s)^2}\left[8+\frac{2}{c(1+c)^4}\cos(kr_sc)\right]+
    \mathcal{O}\left((kr_s)^{-3}\right)\right\}.
\end{equation*}
Unfortunately, in the Bullock, et al. model of halo concentrations, the mean halo concentration vanishes at a maximum halo mass scale. To stay true to the definition of the model, we want to be able to evaluate the transform in the vanishing concentration regime. Here, one should use
\begin{equation*}
  (r_s\rho_s)^2=\left(\frac{\Delta_{\text{vir}}\mean{\rho}R_{\text{vir}}}{3}
    \right)^2\frac{c^4}{\left[\ln(1+c)-\frac{c}{1+c}\right]^2}=
    \left(\frac{\Delta_{\text{vir}}\mean{\rho}R_{\text{vir}}}{3}
    \right)^2\left[4+\frac{32}{3}c+\frac{28}{3}c^2+\mathcal{O}(c^3)\right].
\end{equation*}
If $c\ll1$ and $c\ll kR_{\text{vir}}$ (equivalently, $kr_s\gg1$), then
\begin{eqnarray}
  \nonumber
  k\ \mathcal{FT}\{\rho_h^2\}(k)&=&4\pi(r_s\rho_s)^2\left\{\text{Si}
    (kR_{\text{vir}})-2\Big[1-\cos(kR_{\text{vir}})\Big]\frac{c}
    {kR_{\text{vir}}}+3\Big[\sin(kR_{\text{vir}})-kR_{\text{vir}}
    \cos(kR_{\text{vir}})\Big]\left(\frac{c}{kR_{\text{vir}}}\right)^2\right.\\
  \nonumber
  &&\left.+\mathcal{O}(c^3)+\mathcal{O}\left(\left(\frac{c}{kR_{\text{vir}}}
    \right)^3\right)\right\}
\end{eqnarray}
In the case where $c\ll1$ and $c\ll\!\!\!\!\!\!\!\diagup\ kR_{\text{vir}}$, then we must have $kR_{\text{vir}}\ll1$ and can use
\begin{equation*}
  k\ \mathcal{FT}\{\rho_h^2\}(k)=4\pi(r_s\rho_s)^2\frac{kR_{\text{vir}}}{c}
    \left[\frac{1}{3}\left(1-\frac{1}{(1+c)^3}\right)-\frac{c}{18(1+c)}
    \left(kR_{\text{vir}}\right)^2+\mathcal{O}\left(\left(kR_{\text{vir}}
    \right)^4\right)\right].
\end{equation*}

\subsection{$\mathcal{FT}\{\rho_h^2\sigma_{uh}^2\}(k|M,z)$}

This Fourier transform is simply expressed in the form
\begin{equation}
  \mathcal{FT}\{\rho_h^2\sigma_{uh}^2\}(k)=\frac{4\pi r_s^2\rho_s^2\sigma_s^2}
    {k}\mathcal{S}(kR_{\text{vir}},c)
\end{equation}
where we have defined
\begin{equation}
  \label{eq:Sdef}
  \mathcal{S}(x,c)\equiv\int_0^c\frac{\sin\left(\frac{x}{c}t\right)}
    {t^{1-\beta}(1+t)^q}\der t
\end{equation}
with $\beta=17/27$ as previously defined, and $q=16/3$ for the NFW profile. The important result that allows efficient evaluation of $\mathcal{S}(x,c)$ for a wide range of scales for $x$ and $c$ is the set of expansions (see Appendix~\ref{subap:ftderivation})
\begin{equation}
  \label{eq:den2vvarft}
  \mathcal{S}(x,c)=
  \begin{cases}
    {\displaystyle\frac{c^\beta}{(1+c)^q}\sum_{p=0}^\infty
      \frac{(q)_p}{(\beta)_{p+1}}\Im\Big[
      \mbox{}_1\!F\!_1(\beta;\beta+p+1;ix)\Big]\left(\frac{c}{1+c}\right)^p},
      &{\displaystyle c\leq c_T} \\
    \normalsize\vspace{-0.9\baselineskip}\mbox{}\\
    {\displaystyle\Gamma(\beta)\Im\Big[U\!\left(\beta,\beta-q+1,
      -i\frac{x}{c}\right)\Big]-\frac{c^\beta}{(1+c)^q}\sum_{p=0}^\infty
      (q)_p\Im\Big[e^{ix}U(p+1,\beta-q+1,-ix)\Big]
      \left(\frac{1}{1+c}\right)^p},&{\displaystyle c>c_T}
  \end{cases}
\end{equation}
where $c_T$ is an appropriate transition concentration. The truncation errors of the two expressions were found to be of the same magnitude near $c=0.8$, making it a reasonable value for $c_T$. Also in the expression appears the gamma function $\Gamma(x)$, the Pochhammer symbol
\[(q)_p\equiv\frac{\Gamma(q+p)}{\Gamma(q)}=q(q+1)(q+2)\cdots(q+p-1),\]
the confluent hypergeometric function of the first kind $_1\!F\!_1(a;b;z)$ (expressed in the notation of a generalized hypergeometric function), and the confluent hypergeometric function of the second kind $U(a,b,z)$.

For $c<c_T$, if $x$ is small (say $\alt 4$), then the hypergeometric functions are most efficiently evaluated with their power series
\[
  \Im\Big[\mbox{}_1\!F\!_1(\beta;\beta+p+1;ix)\Big]=\sum_{n=0}^\infty
  \frac{(-1)^n(\beta)_{2n+1}}{(\beta+p+1)_{2n+1}}\frac{x^{2n+1}}{(2n+1)!}.
\]
For larger values of $x$, the functions are quickly determined from the recurrence relation
\[
  _1\!F\!_1(\beta;\beta+p+1;ix)=\frac{\beta+p}{p}\left[\left(1-i\frac{\beta+p
  -1}{x}\right)\mbox{}_1\!F\!_1(\beta;\beta+p;ix)+i\frac{\beta+p-1}{x}\mbox{}
  _1\!F\!_1(\beta;\beta+p-1;ix)\right],
\]
or
\begin{eqnarray}
  \nonumber
  \Re\Big[\mbox{}_1\!F\!_1(\beta;\beta+p+1;ix)\Big]&=&\frac{\beta+p}{p}\left\{
    \Re\Big[\mbox{}_1\!F\!_1(\beta;\beta+p;ix)\Big]+\frac{\beta+p+1}{x}
    \Im\Big[\mbox{}_1\!F\!_1(\beta;\beta+p;ix)-
    \mbox{}_1\!F\!_1(\beta;\beta+p-1;ix)\Big]\right\},\\
  \nonumber
  \Im\Big[\mbox{}_1\!F\!_1(\beta;\beta+p+1;ix)\Big]&=&\frac{\beta+p}{p}\left\{
    \Im\Big[\mbox{}_1\!F\!_1(\beta;\beta+p;ix)\Big]-\frac{\beta+p+1}{x}
    \Re\Big[\mbox{}_1\!F\!_1(\beta;\beta+p;ix)-
    \mbox{}_1\!F\!_1(\beta;\beta+p-1;ix)\Big]\right\}.
\end{eqnarray}
Since $_1\!F\!_1(\beta;\beta;ix)=e^{ix}$, then we only need the numerical evaluation of $_1\!F\!_1(\beta;\beta+1;ix)$ to be able to determine the rest of the sum's hypergeometric functions using the recurrence relation. The power series is suitable for $x\alt10$:
\[
  _1\!F\!_1(\beta;\beta+1;ix)=\sum_{n=0}^\infty\frac{\beta}{\beta+n}
    \frac{(ix)^n}{n!}.
\]
We find the asymptotic expansion converges appropriately for $x\agt27$:
\[
  _1\!F\!_1(\beta;\beta+1;ix)\simeq\Gamma(\beta+1)\exp\!\left(i\frac{\beta\pi}
  {2}\right)x^{-\beta}+\beta\sum_{n=1,2,3,\dots}(n-\beta)_{n-1}\exp\!\left[i\left(x-
  \frac{n\pi}{2}\right)\right]x^{-n}.
\]
For $10\alt x\alt27$, these series' do not converge sufficiently with double machine precision arithmetic. For this short range of $x$, it is not too much of a burden to evaluate the function via numerical integration
\[
  _1\!F\!_1(\beta;\beta+1;ix)=\beta\int_0^1e^{ixt}t^{\beta-1}\der t.
\]

For large concentrations $c>c_T$, there are two components. The first term depends only on the ratio $\bar x\equiv x/c$ and requires the evaluation of $\Im [U(\beta,\beta-q+1,-i\bar x)]$. We use the perturbative expansion for $\bar x\leq5$, for which a convenient expression is
\begin{eqnarray}
  \nonumber
  \Im [U(\beta,\beta-q+1,-i\bar x)]&=&\sum_{n=0}^\infty\left[(-1)^{(n+1)/2}
    (n\ \text{mod}\ 2) \frac{\Gamma(q-\beta)}{\Gamma(q)}\frac{(\beta)_n}
    {(\beta-q+1)_n}\right.\\
  \nonumber
  &&\qquad\left.+\frac{\pi}{2\Gamma(\beta)\Gamma(q-\beta+1)}
      \frac{(-1)^{\lfloor n/2\rfloor}}{\mathcal{CS}_n\!\left(\frac{\pi(q-\beta)}
      {2}\right)}\frac{(q)_n}{(q-\beta+1)_n}\bar x^{q-\beta}\right]\frac{\bar 
      x^n}{n!}
\end{eqnarray}
where we have introduced: the modulo 2 operation
\[
  n\ \text{mod}\ 2=
  \begin{cases}
    0, &\text{$n$ even,}\\
    1, &\text{$n$ odd,}
  \end{cases}
\]
the floor operation $\lfloor x\rfloor$ being the largest integer $\leq x$, and we define the trigonometric function
\[
  \mathcal{CS}_n(x)\equiv
  \begin{cases}
    \cos x, &\text{$n$ even,}\\
    \sin x, &\text{$n$ odd}.
  \end{cases}
\]
The asymptotic expansion
\[
  \Im [U(\beta,\beta-q+1,-i\bar x)]\simeq\bar x^{-\beta}\sum_{n=0,1,2,\dots}
  (-1)^{\lfloor3n/2\rfloor}\mathcal{CS}_{n+1}\!\left(\frac{\beta\pi}{2}\right)
  (\beta)_n(q)_n\frac{\bar x^{-n}}{n!}
\]
works sufficiently for $\bar x\geq40$. For the little remaining range of $5<\bar x<40$, we simply numerically evaluate the integral representation
\[
  \Gamma(\beta)\Im [U(\beta,\beta-q+1,-i\bar x)]=\int_0^\infty\frac{\sin(\bar
  xt)}{t^{1-\beta}(1+t)^q}\der t.
\]
To evaluate the functions in the sum,
\[
  \Im\Big[e^{ix}U(p+1,\beta-q+1,-ix)\Big]=\sin x\ \Re\Big[U(p+1,\beta-q+1,-ix)
  \Big]+\cos x\ \Im\Big[U(p+1,\beta-q+1,-ix)\Big],
\]
we again can make use of recursion relations
\begin{eqnarray}
  \nonumber
  \Re\Big[U(p+1,\beta-q+1,-ix)\Big]&=&\frac{1}{p(p+q-\beta)}\left\{x\Im\Big[
    U(p,\beta-q+1,-ix)\Big]+(2p+q-\beta-1)\Re\Big[U(p,\beta-q+1,-ix)\Big]
    \right.\\
  \nonumber
  &&\left.-\Re\Big[U(p-1,\beta-q+1,-ix)\Big]\right\},\\
  \nonumber
  \Im\Big[U(p+1,\beta-q+1,-ix)\Big]&=&\frac{1}{p(p+q-\beta)}\left\{-x\Re\Big[
    U(p,\beta-q+1,-ix)\Big]+(2p+q-\beta-1)\Im\Big[U(p,\beta-q+1,-ix)\Big]
    \right.\\
  \nonumber
  &&\left.-\Im\Big[U(p-1,\beta-q+1,-ix)\Big]\right\}.
\end{eqnarray}
Since $U(0,\cdot,\cdot)=1$, we only require the evaluation of $U(1,\beta-q+1,-ix)$. For $x\leq4$,
\begin{eqnarray}
  \nonumber
  U(1,\beta-q+1,-ix)&=&\frac{1}{q-\beta}\sum_{n=0}^\infty\left[\frac{(-1)^{n/2}
    (1-n\ \text{mod}\ 2)}{(\beta-q+1)_n}-\frac{\pi x^{q-\beta}}
    {2\Gamma(q-\beta)}\frac{(-1)^{\lfloor(n+1)/2\rfloor}}{n!\mathcal{CS}_{n+1}
    \!\left(\frac{\pi(q-\beta)}{2}\right)}\right]x^n\\
  \nonumber
  &&+\frac{i}{q-\beta}\sum_{n=0}^\infty\left[\frac{(-1)^{(n+1)/2}
    (n\ \text{mod}\ 2)}{(\beta-q+1)_n}-\frac{\pi x^{q-\beta}}
    {2\Gamma(q-\beta)}\frac{(-1)^{\lfloor n/2\rfloor}}{n!\mathcal{CS}_n
    \!\left(\frac{\pi(q-\beta)}{2}\right)}\right]x^n
\end{eqnarray}
was used, and for $x\geq45$, we evaluated
\[
  U(1,\beta-q+1,-ix)\simeq-\sum_{n=0,1,2,\dots}(q-\beta+1)_n(ix)^{-(n+1)}.
\]
For the mid-values of $x$, we numerically integrated
\[
  U(1,\beta-q+1,-ix)=\int_0^\infty\frac{e^{ixt}}{(1+t)^{q-\beta+1}}\der t.
\]
For very large values of $x$, the recursion relations will fail, due to loss of precision from subtracted quantities being very near each other. In this regime, we can evaluate the hypergeometric function in each term of the sum from the asymptotic series
\[
  \Im\Big[e^{ix}U(p+1,\beta-q+1,-ix)\Big]\simeq\sum_{n=0,1,2,\dots} (-1)^{\lfloor
    (n-p)/2\rfloor}\mathcal{CS}_{n+p}\!(x)(p+1)_n(q-\beta+p+1)_n\frac{x^{-(n+p+1)}}
    {n!}.
\]

\subsection{Derivation of Equation~(\ref{eq:den2vvarft})}
\label{subap:ftderivation}

We begin with the case of $c<c_T=0.8$ by expanding $(1+t)^{-q}$ in Equation~(\ref{eq:Sdef}) as a power series, and rescaling $t\longrightarrow xt/c$ to get
\[
  \mathcal{S}(x,c)=\sum_{m=0}^\infty(-1)^m\frac{\Gamma(q+m)}{\Gamma(q)m!}\frac{I_{\beta+m-1}(x)}{x^{\beta+m}}c^{\beta+m}
\]
where
\[
  I_n(x)\equiv\int_0^xt^n\sin t\ \der t.
\]
Then we let $\kappa=c/(1+c)$, write the expression in the form
\[
  \mathcal{S}(x,c)=\frac{c^\beta}{(1+c)^q}\sum_{m=0}^\infty(-1)^m\frac{\Gamma(q+m)}{\Gamma(q)m!}
  \frac{I_{\beta+m-1}(x)}{x^{\beta+m}}\frac{\kappa^m}{(1-\kappa)^{q+m}},
\]
and expand the $\kappa$ expression in a power series with shifted indices
\[
  (1-\kappa)^{-(q+m)}=\sum_{p=0}^\infty\frac{\Gamma(q+m+p)}{\Gamma(q+m)p!}\kappa^p
  =\sum_{p=m}^\infty\frac{\Gamma(q+p)}{\Gamma(q+m)(p-m)!}\kappa^{p-m}.
\]
Swap the order of summation to find
\[
  \mathcal{S}(x,c)=\frac{c^\beta}{(1+c)^q}\sum_{p=0}^\infty\frac{\Gamma(q+p)}{\Gamma(q)}
  \left[\sum_{m=0}^p\frac{(-1)^m}{m!(p-m)!}\frac{I_{\beta+m-1}(x)}{x^{\beta+m}}\right]
  \left(\frac{c}{1+c}\right)^p
\]
after substituting $c$ back into $\kappa$. The quantity in the square brackets can be rewritten using the integral representation of the confluent hypergeometric function
\[
  _1\!F\!_1(a;b;z)=\frac{\Gamma(b)}{\Gamma(a)\Gamma(b-a)}\int_0^1e^{zt}t^{a-1}(1-t)^{b-a-1}\der t
\]
(convergent for $\Re(b)>\Re(a)>0$), giving us
\begin{eqnarray}
  \nonumber
  \sum_{m=0}^p\frac{(-1)^m}{m!(p-m)!}\frac{I_{\beta+m-1}(x)}{x^{\beta+m}}
    &=&\frac{1}{p!}\sum_{m=0}^p(-1)^m\binom{p}{m}x^{-(\beta+m)}\int_0^x t^{\beta+m-1}\sin t\ \der t\\
  \nonumber
  &=&\frac{1}{p!}\int_0^1t^{\beta-1}\left[\sum_{m=0}^p\binom{p}{m}(-t)^m\right]\sin(xt)\ \der t\\
  \nonumber
  &=&\frac{1}{p!}\,\Im\!\left[\int_0^1t^{\beta-1}(1-t)^pe^{ixt}\der t\right]\\
  \nonumber
  &=&\frac{\Gamma(\beta)}{\Gamma(\beta+p+1)}\Im\Big[\mbox{}_1\!F\!_1(\beta;\beta+p+1;ix)\Big]
\end{eqnarray}
where, in the second line, we rescaled $t\longrightarrow t/x$.

For the case of large concentrations $c>c_T$, we break $\mathcal{S}$ into two terms
\[
  \mathcal{S}(x,c)=\mathcal{S}_1\!\left(\frac{x}{c}\right)-\mathcal{S}_2(x,c)
\]
with
\begin{eqnarray}
  \nonumber
  \mathcal{S}_1(\bar{x})&=&\int_0^\infty\frac{\sin(\bar{x}t)}{t^{1-\beta}(1+t)^q}\der t,\\
  \nonumber
  \mathcal{S}_2(x,c)&=&\int_c^\infty\frac{\sin\left(\frac{x}{c}t\right)}
    {t^{1-\beta}(1+t)^q}\der t.
\end{eqnarray}
The first term is simply
\[
  \mathcal{S}_1(\bar{x})=\Im\left[\int_0^\infty\frac{e^{i\bar{x}t}}{t^{1-\beta}(1+t)^q}\der t\right]
  =\Gamma(\beta)\Im\Big[U(\beta,\beta-q+1,-i\bar{x})\Big]
\]
given that the confluent hypergeometric function of the second kind has the integral representation
\[
  U(a,b,z)=\frac{1}{\Gamma(a)}\int_0^\infty e^{-zt}t^{a-1}(1+t)^{b-a-1}\der t
\]
if $\Re(a)>0$ and $\Re(z)>0$. For the second term, first substitute $t\longrightarrow t-c$
\begin{eqnarray}
  \nonumber
  \mathcal{S}_2(x,c)&=&\int_0^\infty\frac{\sin\left(\frac{x}{c}(c+t)\right)}
    {(c+t)^{1-\beta}(1+c+t)^q}\der t\\
  \nonumber
  &=&\frac{c^\beta}{(1+c)^q}\int_0^\infty\frac{\sin\left[x\left(1+\frac{t}{c}\right)\right]}
    {\left(1+\frac{t}{c}\right)^{1-\beta}\left(1+\frac{t}{1+c}\right)^q}\frac{\der t}{c},
\end{eqnarray}
and then substitute $t\longrightarrow t/c$
\[
  \mathcal{S}_2(x,c)=\frac{c^\beta}{(1+c)^q}\,\Im\!\left[e^{ix}\int_0^\infty\frac{e^{ixt}}
    {(1+t)^{1-\beta}(1+\kappa t)^q}\der t\right]
\]
with $\kappa=c/(1+c)$, as before. As is appropriate for large values of $c$, we expand as a power series about $\kappa=1$.
\begin{eqnarray}
  \nonumber
  \mathcal{S}_2(x,c)&=&\frac{c^\beta}{(1+c)^q}\sum_{p=0}^\infty\frac{\Gamma(q+p)}{\Gamma(q)}
    \Im\Big[e^{ix}\int_0^\infty e^{ixt}t^p(1+t)^{\beta-q-p-1}\der t\Big]\frac{(1-\kappa)^p}{p!}\\
  \nonumber
  &=&\frac{c^\beta}{(1+c)^q}\sum_{p=0}^\infty\frac{\Gamma(q+p)}{\Gamma(q)}
    \Im\Big[e^{ix}U(p+1,\beta-q+1,-ix)\Big]\left(\frac{1}{1+c}\right)^p
\end{eqnarray}

\bibliography{pwavepowspec}

\begin{thebibliography}{49}%
\makeatletter
\providecommand \@ifxundefined [1]{%
 \@ifx{#1\undefined}
}%
\providecommand \@ifnum [1]{%
 \ifnum #1\expandafter \@firstoftwo
 \else \expandafter \@secondoftwo
 \fi
}%
\providecommand \@ifx [1]{%
 \ifx #1\expandafter \@firstoftwo
 \else \expandafter \@secondoftwo
 \fi
}%
\providecommand \natexlab [1]{#1}%
\providecommand \enquote  [1]{``#1''}%
\providecommand \bibnamefont  [1]{#1}%
\providecommand \bibfnamefont [1]{#1}%
\providecommand \citenamefont [1]{#1}%
\providecommand \href@noop [0]{\@secondoftwo}%
\providecommand \href [0]{\begingroup \@sanitize@url \@href}%
\providecommand \@href[1]{\@@startlink{#1}\@@href}%
\providecommand \@@href[1]{\endgroup#1\@@endlink}%
\providecommand \@sanitize@url [0]{\catcode `\\12\catcode `\$12\catcode
  `\&12\catcode `\#12\catcode `\^12\catcode `\_12\catcode `\%12\relax}%
\providecommand \@@startlink[1]{}%
\providecommand \@@endlink[0]{}%
\providecommand \url  [0]{\begingroup\@sanitize@url \@url }%
\providecommand \@url [1]{\endgroup\@href {#1}{\urlprefix }}%
\providecommand \urlprefix  [0]{URL }%
\providecommand \Eprint [0]{\href }%
\@ifxundefined \urlstyle {%
  \providecommand \doi  [0]{\begingroup \@sanitize@url \@doi}%
  \providecommand \@doi [1]{\endgroup \@@startlink {\doibase
  #1}doi:\discretionary {}{}{}#1\@@endlink }%
}{%
  \providecommand \doi  [0]{doi:\discretionary{}{}{}\begingroup
  \urlstyle{rm}\Url }%
}%
\providecommand \doibase [0]{http://dx.doi.org/}%
\providecommand \Doi [0]{\begingroup \@sanitize@url \@Doi }%
\providecommand \@Doi  [1]{\endgroup\@@startlink{\doibase#1}\@@Doi}%
\providecommand \@@Doi [1]{#1\@@endlink}%
\providecommand \selectlanguage [0]{\@gobble}%
\providecommand \bibinfo  [0]{\@secondoftwo}%
\providecommand \bibfield  [0]{\@secondoftwo}%
\providecommand \translation [1]{[#1]}%
\providecommand \BibitemOpen [0]{}%
\providecommand \bibitemStop [0]{}%
\providecommand \bibitemNoStop [0]{.\EOS\space}%
\providecommand \EOS [0]{\spacefactor3000\relax}%
\providecommand \BibitemShut  [1]{\csname bibitem#1\endcsname}%
\bibitem [{\citenamefont {Komatsu}\ \emph {et~al.}(2009)\citenamefont {Komatsu}
  \emph {et~al.}}]{Komatsu:2008hk}%
  \BibitemOpen
  \bibfield  {author} {\bibinfo {author} {\bibfnamefont {E.}~\bibnamefont
  {Komatsu}} \emph {et~al.} (\bibinfo {collaboration} {WMAP}),\ }\Doi
  {10.1088/0067-0049/180/2/330} {\bibfield  {journal} {\bibinfo  {journal}
  {Astrophys. J. Suppl.},\ }\textbf {\bibinfo {volume} {180}},\ \bibinfo
  {pages} {330} (\bibinfo {year} {2009})},\ \Eprint
  {http://arxiv.org/abs/0803.0547} {arXiv:0803.0547 [astro-ph]} \BibitemShut
  {NoStop}%
\bibitem [{\citenamefont {Komatsu}\ \emph {et~al.}(2011)\citenamefont {Komatsu}
  \emph {et~al.}}]{Komatsu:2010fb}%
  \BibitemOpen
  \bibfield  {author} {\bibinfo {author} {\bibfnamefont {E.}~\bibnamefont
  {Komatsu}} \emph {et~al.} (\bibinfo {collaboration} {WMAP}),\ }\Doi
  {10.1088/0067-0049/192/2/18} {\bibfield  {journal} {\bibinfo  {journal}
  {Astrophys. J. Suppl.},\ }\textbf {\bibinfo {volume} {192}},\ \bibinfo
  {pages} {18} (\bibinfo {year} {2011})},\ \Eprint
  {http://arxiv.org/abs/1001.4538} {arXiv:1001.4538 [astro-ph.CO]} \BibitemShut
  {NoStop}%
\bibitem [{\citenamefont {Porter}\ \emph {et~al.}(2011)\citenamefont {Porter},
  \citenamefont {Johnson},\ and\ \citenamefont {Graham}}]{Porter:2011nv}%
  \BibitemOpen
  \bibfield  {author} {\bibinfo {author} {\bibfnamefont {T.~A.}\ \bibnamefont
  {Porter}}, \bibinfo {author} {\bibfnamefont {R.~P.}\ \bibnamefont {Johnson}},
  \ and\ \bibinfo {author} {\bibfnamefont {P.~W.}\ \bibnamefont {Graham}},\
  }\href@noop {} {\bibfield  {journal} {\bibinfo  {journal} {to appear in
  Annual Reviews of Astronomy and Astrophysics}} (\bibinfo {year} {2011})},\
  \Eprint {http://arxiv.org/abs/1104.2836} {arXiv:1104.2836 [astro-ph.HE]}
  \BibitemShut {NoStop}%
\bibitem [{\citenamefont {Adriani}\ \emph {et~al.}(2009)\citenamefont {Adriani}
  \emph {et~al.}}]{Adriani:2008zr}%
  \BibitemOpen
  \bibfield  {author} {\bibinfo {author} {\bibfnamefont {O.}~\bibnamefont
  {Adriani}} \emph {et~al.} (\bibinfo {collaboration} {PAMELA}),\ }\Doi
  {10.1038/nature07942} {\bibfield  {journal} {\bibinfo  {journal} {Nature},\
  }\textbf {\bibinfo {volume} {458}},\ \bibinfo {pages} {607} (\bibinfo {year}
  {2009})},\ \Eprint {http://arxiv.org/abs/0810.4995} {arXiv:0810.4995
  [astro-ph]} \BibitemShut {NoStop}%
\bibitem [{\citenamefont {Abdo}\ \emph {et~al.}(2009)\citenamefont {Abdo} \emph
  {et~al.}}]{Abdo:2009zk}%
  \BibitemOpen
  \bibfield  {author} {\bibinfo {author} {\bibfnamefont {A.~A.}\ \bibnamefont
  {Abdo}} \emph {et~al.} (\bibinfo {collaboration} {The Fermi LAT}),\ }\Doi
  {10.1103/PhysRevLett.102.181101} {\bibfield  {journal} {\bibinfo  {journal}
  {Phys. Rev. Lett.},\ }\textbf {\bibinfo {volume} {102}},\ \bibinfo {pages}
  {181101} (\bibinfo {year} {2009})},\ \Eprint {http://arxiv.org/abs/0905.0025}
  {arXiv:0905.0025 [astro-ph.HE]} \BibitemShut {NoStop}%
\bibitem [{\citenamefont {Ackermann}\ \emph {et~al.}(2010)\citenamefont
  {Ackermann} \emph {et~al.}}]{Ackermann:2010ij}%
  \BibitemOpen
  \bibfield  {author} {\bibinfo {author} {\bibfnamefont {M.}~\bibnamefont
  {Ackermann}} \emph {et~al.} (\bibinfo {collaboration} {Fermi LAT}),\
  }\href@noop {} { (\bibinfo {year} {2010})},\ \Eprint
  {http://arxiv.org/abs/1008.3999} {arXiv:1008.3999 [astro-ph.HE]} \BibitemShut
  {NoStop}%
\bibitem [{\citenamefont {Biermann}\ \emph {et~al.}(2009)\citenamefont
  {Biermann} \emph {et~al.}}]{Biermann:2009qi}%
  \BibitemOpen
  \bibfield  {author} {\bibinfo {author} {\bibfnamefont {P.~L.}\ \bibnamefont
  {Biermann}} \emph {et~al.},\ }\Doi {10.1103/PhysRevLett.103.061101}
  {\bibfield  {journal} {\bibinfo  {journal} {Phys. Rev. Lett.},\ }\textbf
  {\bibinfo {volume} {103}},\ \bibinfo {pages} {061101} (\bibinfo {year}
  {2009})},\ \Eprint {http://arxiv.org/abs/0903.4048} {arXiv:0903.4048
  [astro-ph.HE]} \BibitemShut {NoStop}%
\bibitem [{\citenamefont {Profumo}(2008)}]{Profumo:2008ms}%
  \BibitemOpen
  \bibfield  {author} {\bibinfo {author} {\bibfnamefont {S.}~\bibnamefont
  {Profumo}},\ }\href@noop {} { (\bibinfo {year} {2008})},\ \Eprint
  {http://arxiv.org/abs/0812.4457} {arXiv:0812.4457 [astro-ph]} \BibitemShut
  {NoStop}%
\bibitem [{\citenamefont {Grasso}\ \emph {et~al.}(2009)\citenamefont {Grasso}
  \emph {et~al.}}]{Grasso:2009ma}%
  \BibitemOpen
  \bibfield  {author} {\bibinfo {author} {\bibfnamefont {D.}~\bibnamefont
  {Grasso}} \emph {et~al.} (\bibinfo {collaboration} {FERMI-LAT}),\ }\Doi
  {10.1016/j.astropartphys.2009.07.003} {\bibfield  {journal} {\bibinfo
  {journal} {Astropart. Phys.},\ }\textbf {\bibinfo {volume} {32}},\ \bibinfo
  {pages} {140} (\bibinfo {year} {2009})},\ \Eprint
  {http://arxiv.org/abs/0905.0636} {arXiv:0905.0636 [astro-ph.HE]} \BibitemShut
  {NoStop}%
\bibitem [{\citenamefont {Kamae}\ \emph {et~al.}(2010)\citenamefont {Kamae}
  \emph {et~al.}}]{Kamae:2010ad}%
  \BibitemOpen
  \bibfield  {author} {\bibinfo {author} {\bibfnamefont {T.}~\bibnamefont
  {Kamae}} \emph {et~al.},\ }\href@noop {} { (\bibinfo {year} {2010})},\
  \Eprint {http://arxiv.org/abs/1010.3477} {arXiv:1010.3477 [astro-ph.HE]}
  \BibitemShut {NoStop}%
\bibitem [{\citenamefont {Ando}\ and\ \citenamefont
  {Komatsu}(2006)}]{Ando:2005xg}%
  \BibitemOpen
  \bibfield  {author} {\bibinfo {author} {\bibfnamefont {S.}~\bibnamefont
  {Ando}}\ and\ \bibinfo {author} {\bibfnamefont {E.}~\bibnamefont {Komatsu}},\
  }\Doi {10.1103/PhysRevD.73.023521} {\bibfield  {journal} {\bibinfo  {journal}
  {Phys. Rev.},\ }\textbf {\bibinfo {volume} {D73}},\ \bibinfo {pages} {023521}
  (\bibinfo {year} {2006})},\ \Eprint {http://arxiv.org/abs/astro-ph/0512217}
  {arXiv:astro-ph/0512217} \BibitemShut {NoStop}%
\bibitem [{\citenamefont {Ando}\ \emph {et~al.}(2007)\citenamefont {Ando},
  \citenamefont {Komatsu}, \citenamefont {Narumoto},\ and\ \citenamefont
  {Totani}}]{Ando:2006cr}%
  \BibitemOpen
  \bibfield  {author} {\bibinfo {author} {\bibfnamefont {S.}~\bibnamefont
  {Ando}}, \bibinfo {author} {\bibfnamefont {E.}~\bibnamefont {Komatsu}},
  \bibinfo {author} {\bibfnamefont {T.}~\bibnamefont {Narumoto}}, \ and\
  \bibinfo {author} {\bibfnamefont {T.}~\bibnamefont {Totani}},\ }\Doi
  {10.1103/PhysRevD.75.063519} {\bibfield  {journal} {\bibinfo  {journal}
  {Phys. Rev.},\ }\textbf {\bibinfo {volume} {D75}},\ \bibinfo {pages} {063519}
  (\bibinfo {year} {2007})},\ \Eprint {http://arxiv.org/abs/astro-ph/0612467}
  {arXiv:astro-ph/0612467} \BibitemShut {NoStop}%
\bibitem [{\citenamefont {Cuoco}\ \emph {et~al.}(2007)\citenamefont {Cuoco},
  \citenamefont {Brandbyge}, \citenamefont {Hannestad}, \citenamefont
  {Haugboelle},\ and\ \citenamefont {Miele}}]{Cuoco:2007sh}%
  \BibitemOpen
  \bibfield  {author} {\bibinfo {author} {\bibfnamefont {A.}~\bibnamefont
  {Cuoco}}, \bibinfo {author} {\bibfnamefont {J.}~\bibnamefont {Brandbyge}},
  \bibinfo {author} {\bibfnamefont {S.}~\bibnamefont {Hannestad}}, \bibinfo
  {author} {\bibfnamefont {T.}~\bibnamefont {Haugboelle}}, \ and\ \bibinfo
  {author} {\bibfnamefont {G.}~\bibnamefont {Miele}},\ }\href@noop {} {
  (\bibinfo {year} {2007})},\ \Eprint {http://arxiv.org/abs/0710.4136}
  {arXiv:0710.4136 [astro-ph]} \BibitemShut {NoStop}%
\bibitem [{\citenamefont {Siegal-Gaskins}(2008)}]{SiegalGaskins:2008ge}%
  \BibitemOpen
  \bibfield  {author} {\bibinfo {author} {\bibfnamefont {J.~M.}\ \bibnamefont
  {Siegal-Gaskins}},\ }\Doi {10.1088/1475-7516/2008/10/040} {\bibfield
  {journal} {\bibinfo  {journal} {JCAP},\ }\textbf {\bibinfo {volume} {0810}},\
  \bibinfo {pages} {040} (\bibinfo {year} {2008})},\ \Eprint
  {http://arxiv.org/abs/0807.1328} {arXiv:0807.1328 [astro-ph]} \BibitemShut
  {NoStop}%
\bibitem [{\citenamefont {Campbell}\ \emph {et~al.}(2010)\citenamefont
  {Campbell}, \citenamefont {Dutta},\ and\ \citenamefont
  {Komatsu}}]{Campbell:2010xc}%
  \BibitemOpen
  \bibfield  {author} {\bibinfo {author} {\bibfnamefont {S.}~\bibnamefont
  {Campbell}}, \bibinfo {author} {\bibfnamefont {B.}~\bibnamefont {Dutta}}, \
  and\ \bibinfo {author} {\bibfnamefont {E.}~\bibnamefont {Komatsu}},\ }\Doi
  {10.1103/PhysRevD.82.095007} {\bibfield  {journal} {\bibinfo  {journal}
  {Phys. Rev.},\ }\textbf {\bibinfo {volume} {D82}},\ \bibinfo {pages} {095007}
  (\bibinfo {year} {2010})},\ \Eprint {http://arxiv.org/abs/1009.3530}
  {arXiv:1009.3530 [hep-ph]} \BibitemShut {NoStop}%
\bibitem [{\citenamefont {Drees}\ and\ \citenamefont
  {Nojiri}(1993)}]{Drees:1992am}%
  \BibitemOpen
  \bibfield  {author} {\bibinfo {author} {\bibfnamefont {M.}~\bibnamefont
  {Drees}}\ and\ \bibinfo {author} {\bibfnamefont {M.~M.}\ \bibnamefont
  {Nojiri}},\ }\Doi {10.1103/PhysRevD.47.376} {\bibfield  {journal} {\bibinfo
  {journal} {Phys. Rev.},\ }\textbf {\bibinfo {volume} {D47}},\ \bibinfo
  {pages} {376} (\bibinfo {year} {1993})},\ \Eprint
  {http://arxiv.org/abs/hep-ph/9207234} {arXiv:hep-ph/9207234} \BibitemShut
  {NoStop}%
\bibitem [{\citenamefont {Hisano}\ \emph {et~al.}(2005)\citenamefont {Hisano},
  \citenamefont {Matsumoto}, \citenamefont {Nojiri},\ and\ \citenamefont
  {Saito}}]{Hisano:2004ds}%
  \BibitemOpen
  \bibfield  {author} {\bibinfo {author} {\bibfnamefont {J.}~\bibnamefont
  {Hisano}}, \bibinfo {author} {\bibfnamefont {S.}~\bibnamefont {Matsumoto}},
  \bibinfo {author} {\bibfnamefont {M.~M.}\ \bibnamefont {Nojiri}}, \ and\
  \bibinfo {author} {\bibfnamefont {O.}~\bibnamefont {Saito}},\ }\Doi
  {10.1103/PhysRevD.71.063528} {\bibfield  {journal} {\bibinfo  {journal}
  {Phys. Rev.},\ }\textbf {\bibinfo {volume} {D71}},\ \bibinfo {pages} {063528}
  (\bibinfo {year} {2005})},\ \Eprint {http://arxiv.org/abs/hep-ph/0412403}
  {arXiv:hep-ph/0412403} \BibitemShut {NoStop}%
\bibitem [{\citenamefont {Arkani-Hamed}\ \emph {et~al.}(2009)\citenamefont
  {Arkani-Hamed}, \citenamefont {Finkbeiner}, \citenamefont {Slatyer},\ and\
  \citenamefont {Weiner}}]{ArkaniHamed:2008qn}%
  \BibitemOpen
  \bibfield  {author} {\bibinfo {author} {\bibfnamefont {N.}~\bibnamefont
  {Arkani-Hamed}}, \bibinfo {author} {\bibfnamefont {D.~P.}\ \bibnamefont
  {Finkbeiner}}, \bibinfo {author} {\bibfnamefont {T.~R.}\ \bibnamefont
  {Slatyer}}, \ and\ \bibinfo {author} {\bibfnamefont {N.}~\bibnamefont
  {Weiner}},\ }\Doi {10.1103/PhysRevD.79.015014} {\bibfield  {journal}
  {\bibinfo  {journal} {Phys. Rev.},\ }\textbf {\bibinfo {volume} {D79}},\
  \bibinfo {pages} {015014} (\bibinfo {year} {2009})},\ \Eprint
  {http://arxiv.org/abs/0810.0713} {arXiv:0810.0713 [hep-ph]} \BibitemShut
  {NoStop}%
\bibitem [{\citenamefont {Lattanzi}\ and\ \citenamefont
  {Silk}(2009)}]{Lattanzi:2008qa}%
  \BibitemOpen
  \bibfield  {author} {\bibinfo {author} {\bibfnamefont {M.}~\bibnamefont
  {Lattanzi}}\ and\ \bibinfo {author} {\bibfnamefont {J.~I.}\ \bibnamefont
  {Silk}},\ }\Doi {10.1103/PhysRevD.79.083523} {\bibfield  {journal} {\bibinfo
  {journal} {Phys. Rev.},\ }\textbf {\bibinfo {volume} {D79}},\ \bibinfo
  {pages} {083523} (\bibinfo {year} {2009})},\ \Eprint
  {http://arxiv.org/abs/0812.0360} {arXiv:0812.0360 [astro-ph]} \BibitemShut
  {NoStop}%
\bibitem [{\citenamefont {March-Russell}\ and\ \citenamefont
  {West}(2009)}]{MarchRussell:2008tu}%
  \BibitemOpen
  \bibfield  {author} {\bibinfo {author} {\bibfnamefont {J.~D.}\ \bibnamefont
  {March-Russell}}\ and\ \bibinfo {author} {\bibfnamefont {S.~M.}\ \bibnamefont
  {West}},\ }\Doi {10.1016/j.physletb.2009.04.010} {\bibfield  {journal}
  {\bibinfo  {journal} {Phys. Lett.},\ }\textbf {\bibinfo {volume} {B676}},\
  \bibinfo {pages} {133} (\bibinfo {year} {2009})},\ \Eprint
  {http://arxiv.org/abs/0812.0559} {arXiv:0812.0559 [astro-ph]} \BibitemShut
  {NoStop}%
\bibitem [{\citenamefont {Ibe}\ \emph {et~al.}(2009)\citenamefont {Ibe},
  \citenamefont {Murayama},\ and\ \citenamefont {Yanagida}}]{Ibe:2008ye}%
  \BibitemOpen
  \bibfield  {author} {\bibinfo {author} {\bibfnamefont {M.}~\bibnamefont
  {Ibe}}, \bibinfo {author} {\bibfnamefont {H.}~\bibnamefont {Murayama}}, \
  and\ \bibinfo {author} {\bibfnamefont {T.~T.}\ \bibnamefont {Yanagida}},\
  }\Doi {10.1103/PhysRevD.79.095009} {\bibfield  {journal} {\bibinfo  {journal}
  {Phys. Rev.},\ }\textbf {\bibinfo {volume} {D79}},\ \bibinfo {pages} {095009}
  (\bibinfo {year} {2009})},\ \Eprint {http://arxiv.org/abs/0812.0072}
  {arXiv:0812.0072 [hep-ph]} \BibitemShut {NoStop}%
\bibitem [{\citenamefont {Feldman}\ \emph {et~al.}(2009)\citenamefont
  {Feldman}, \citenamefont {Liu},\ and\ \citenamefont {Nath}}]{Feldman:2008xs}%
  \BibitemOpen
  \bibfield  {author} {\bibinfo {author} {\bibfnamefont {D.}~\bibnamefont
  {Feldman}}, \bibinfo {author} {\bibfnamefont {Z.}~\bibnamefont {Liu}}, \ and\
  \bibinfo {author} {\bibfnamefont {P.}~\bibnamefont {Nath}},\ }\Doi
  {10.1103/PhysRevD.79.063509} {\bibfield  {journal} {\bibinfo  {journal}
  {Phys. Rev.},\ }\textbf {\bibinfo {volume} {D79}},\ \bibinfo {pages} {063509}
  (\bibinfo {year} {2009})},\ \Eprint {http://arxiv.org/abs/0810.5762}
  {arXiv:0810.5762 [hep-ph]} \BibitemShut {NoStop}%
\bibitem [{\citenamefont {Afshordi}\ \emph {et~al.}(2010)\citenamefont
  {Afshordi}, \citenamefont {Mohayaee},\ and\ \citenamefont
  {Bertschinger}}]{Afshordi:2009hn}%
  \BibitemOpen
  \bibfield  {author} {\bibinfo {author} {\bibfnamefont {N.}~\bibnamefont
  {Afshordi}}, \bibinfo {author} {\bibfnamefont {R.}~\bibnamefont {Mohayaee}},
  \ and\ \bibinfo {author} {\bibfnamefont {E.}~\bibnamefont {Bertschinger}},\
  }\Doi {10.1103/PhysRevD.81.101301} {\bibfield  {journal} {\bibinfo  {journal}
  {Phys. Rev.},\ }\textbf {\bibinfo {volume} {D81}},\ \bibinfo {pages} {101301}
  (\bibinfo {year} {2010})},\ \Eprint {http://arxiv.org/abs/0911.0414}
  {arXiv:0911.0414 [astro-ph.CO]} \BibitemShut {NoStop}%
\bibitem [{\citenamefont {Navarro}\ \emph {et~al.}(2008)\citenamefont {Navarro}
  \emph {et~al.}}]{Navarro:2008kc}%
  \BibitemOpen
  \bibfield  {author} {\bibinfo {author} {\bibfnamefont {J.~F.}\ \bibnamefont
  {Navarro}} \emph {et~al.},\ }\href@noop {} { (\bibinfo {year} {2008})},\
  \Eprint {http://arxiv.org/abs/0810.1522} {arXiv:0810.1522 [astro-ph]}
  \BibitemShut {NoStop}%
\bibitem [{Note1()}]{Note1}%
  \BibitemOpen
  \bibinfo {note} {The assumptions of the velocity distribution may easily be
  relaxed to more general cases in future work.}\BibitemShut {Stop}%
\bibitem [{\citenamefont {Stecker}\ \emph {et~al.}(2006)\citenamefont
  {Stecker}, \citenamefont {Malkan},\ and\ \citenamefont
  {Scully}}]{Stecker:2005qs}%
  \BibitemOpen
  \bibfield  {author} {\bibinfo {author} {\bibfnamefont {F.~W.}\ \bibnamefont
  {Stecker}}, \bibinfo {author} {\bibfnamefont {M.~A.}\ \bibnamefont {Malkan}},
  \ and\ \bibinfo {author} {\bibfnamefont {S.~T.}\ \bibnamefont {Scully}},\
  }\Doi {10.1086/506188} {\bibfield  {journal} {\bibinfo  {journal} {Astrophys.
  J.},\ }\textbf {\bibinfo {volume} {648}},\ \bibinfo {pages} {774} (\bibinfo
  {year} {2006})},\ \Eprint {http://arxiv.org/abs/astro-ph/0510449}
  {arXiv:astro-ph/0510449} \BibitemShut {NoStop}%
\bibitem [{\citenamefont {Stecker}\ \emph {et~al.}(2007)\citenamefont
  {Stecker}, \citenamefont {Malkan},\ and\ \citenamefont
  {Scully}}]{Stecker:2006eh}%
  \BibitemOpen
  \bibfield  {author} {\bibinfo {author} {\bibfnamefont {F.~W.}\ \bibnamefont
  {Stecker}}, \bibinfo {author} {\bibfnamefont {M.~A.}\ \bibnamefont {Malkan}},
  \ and\ \bibinfo {author} {\bibfnamefont {S.~T.}\ \bibnamefont {Scully}},\
  }\Doi {10.1086/511738} {\bibfield  {journal} {\bibinfo  {journal} {Astrophys.
  J.},\ }\textbf {\bibinfo {volume} {658}},\ \bibinfo {pages} {1392} (\bibinfo
  {year} {2007})},\ \Eprint {http://arxiv.org/abs/astro-ph/0612048}
  {arXiv:astro-ph/0612048} \BibitemShut {NoStop}%
\bibitem [{\citenamefont {Gondolo}\ \emph {et~al.}(2004)\citenamefont {Gondolo}
  \emph {et~al.}}]{Gondolo:2004sc}%
  \BibitemOpen
  \bibfield  {author} {\bibinfo {author} {\bibfnamefont {P.}~\bibnamefont
  {Gondolo}} \emph {et~al.},\ }\Doi {10.1088/1475-7516/2004/07/008} {\bibfield
  {journal} {\bibinfo  {journal} {JCAP},\ }\textbf {\bibinfo {volume} {0407}},\
  \bibinfo {pages} {008} (\bibinfo {year} {2004})},\ \Eprint
  {http://arxiv.org/abs/astro-ph/0406204} {arXiv:astro-ph/0406204} \BibitemShut
  {NoStop}%
\bibitem [{\citenamefont {Paige}\ \emph {et~al.}(2003)\citenamefont {Paige},
  \citenamefont {Protopopescu}, \citenamefont {Baer},\ and\ \citenamefont
  {Tata}}]{Paige:2003mg}%
  \BibitemOpen
  \bibfield  {author} {\bibinfo {author} {\bibfnamefont {F.~E.}\ \bibnamefont
  {Paige}}, \bibinfo {author} {\bibfnamefont {S.~D.}\ \bibnamefont
  {Protopopescu}}, \bibinfo {author} {\bibfnamefont {H.}~\bibnamefont {Baer}},
  \ and\ \bibinfo {author} {\bibfnamefont {X.}~\bibnamefont {Tata}},\
  }\href@noop {} { (\bibinfo {year} {2003})},\ \Eprint
  {http://arxiv.org/abs/hep-ph/0312045} {arXiv:hep-ph/0312045} \BibitemShut
  {NoStop}%
\bibitem [{\citenamefont {Hahn}\ \emph {et~al.}(2009)\citenamefont {Hahn},
  \citenamefont {Heinemeyer}, \citenamefont {Hollik}, \citenamefont {Rzehak},\
  and\ \citenamefont {Weiglein}}]{Hahn:2009zz}%
  \BibitemOpen
  \bibfield  {author} {\bibinfo {author} {\bibfnamefont {T.}~\bibnamefont
  {Hahn}}, \bibinfo {author} {\bibfnamefont {S.}~\bibnamefont {Heinemeyer}},
  \bibinfo {author} {\bibfnamefont {W.}~\bibnamefont {Hollik}}, \bibinfo
  {author} {\bibfnamefont {H.}~\bibnamefont {Rzehak}}, \ and\ \bibinfo {author}
  {\bibfnamefont {G.}~\bibnamefont {Weiglein}},\ }\Doi
  {10.1016/j.cpc.2009.02.014} {\bibfield  {journal} {\bibinfo  {journal}
  {Comput. Phys. Commun.},\ }\textbf {\bibinfo {volume} {180}},\ \bibinfo
  {pages} {1426} (\bibinfo {year} {2009})}\BibitemShut {NoStop}%
\bibitem [{Note2()}]{Note2}%
  \BibitemOpen
  \bibinfo {note} {When describing $\protect \mathaccentV {tilde}07E{\xi }_h$,
  we refer to it as the \protect \emph {full} halo correlation function to
  differentiate it from $\xi _h$.}\BibitemShut {Stop}%
\bibitem [{\citenamefont {Tseliakhovich}\ and\ \citenamefont
  {Hirata}(2010)}]{Tseliakhovich:2010bj}%
  \BibitemOpen
  \bibfield  {author} {\bibinfo {author} {\bibfnamefont {D.}~\bibnamefont
  {Tseliakhovich}}\ and\ \bibinfo {author} {\bibfnamefont {C.}~\bibnamefont
  {Hirata}},\ }\Doi {10.1103/PhysRevD.82.083520} {\bibfield  {journal}
  {\bibinfo  {journal} {Phys. Rev.},\ }\textbf {\bibinfo {volume} {D82}},\
  \bibinfo {pages} {083520} (\bibinfo {year} {2010})},\ \Eprint
  {http://arxiv.org/abs/1005.2416} {arXiv:1005.2416 [astro-ph.CO]} \BibitemShut
  {NoStop}%
\bibitem [{\citenamefont {Shandera}\ \emph {et~al.}(2011)\citenamefont
  {Shandera}, \citenamefont {Dalal},\ and\ \citenamefont
  {Huterer}}]{Shandera:2010ei}%
  \BibitemOpen
  \bibfield  {author} {\bibinfo {author} {\bibfnamefont {S.}~\bibnamefont
  {Shandera}}, \bibinfo {author} {\bibfnamefont {N.}~\bibnamefont {Dalal}}, \
  and\ \bibinfo {author} {\bibfnamefont {D.}~\bibnamefont {Huterer}},\ }\Doi
  {10.1088/1475-7516/2011/03/017} {\bibfield  {journal} {\bibinfo  {journal}
  {JCAP},\ }\textbf {\bibinfo {volume} {1103}},\ \bibinfo {pages} {017}
  (\bibinfo {year} {2011})},\ \Eprint {http://arxiv.org/abs/1010.3722}
  {arXiv:1010.3722 [astro-ph.CO]} \BibitemShut {NoStop}%
\bibitem [{\citenamefont {Parfrey}\ \emph {et~al.}(2011)\citenamefont
  {Parfrey}, \citenamefont {Hui},\ and\ \citenamefont
  {Sheth}}]{Parfrey:2010uy}%
  \BibitemOpen
  \bibfield  {author} {\bibinfo {author} {\bibfnamefont {K.}~\bibnamefont
  {Parfrey}}, \bibinfo {author} {\bibfnamefont {L.}~\bibnamefont {Hui}}, \ and\
  \bibinfo {author} {\bibfnamefont {R.~K.}\ \bibnamefont {Sheth}},\ }\Doi
  {10.1103/PhysRevD.83.063511} {\bibfield  {journal} {\bibinfo  {journal}
  {Phys. Rev.},\ }\textbf {\bibinfo {volume} {D83}},\ \bibinfo {pages} {063511}
  (\bibinfo {year} {2011})},\ \Eprint {http://arxiv.org/abs/1012.1335}
  {arXiv:1012.1335 [astro-ph.CO]} \BibitemShut {NoStop}%
\bibitem [{\citenamefont {Peebles}(1980)}]{Peebles:1980}%
  \BibitemOpen
  \bibfield  {author} {\bibinfo {author} {\bibfnamefont {P.~J.~E.}\
  \bibnamefont {Peebles}},\ }\href@noop {} {\emph {\bibinfo {title} {{The
  Large-Scale Structure of the Universe}}}}\ (\bibinfo  {publisher} {Princeton
  University Press},\ \bibinfo {year} {1980})\BibitemShut {NoStop}%
\bibitem [{\citenamefont {Sheth}\ and\ \citenamefont
  {Tormen}(2002)}]{Sheth:2001dp}%
  \BibitemOpen
  \bibfield  {author} {\bibinfo {author} {\bibfnamefont {R.~K.}\ \bibnamefont
  {Sheth}}\ and\ \bibinfo {author} {\bibfnamefont {G.}~\bibnamefont {Tormen}},\
  }\Doi {10.1046/j.1365-8711.2002.04950.x} {\bibfield  {journal} {\bibinfo
  {journal} {Mon. Not. Roy. Astron. Soc.},\ }\textbf {\bibinfo {volume}
  {329}},\ \bibinfo {pages} {61} (\bibinfo {year} {2002})},\ \Eprint
  {http://arxiv.org/abs/astro-ph/0105113} {arXiv:astro-ph/0105113} \BibitemShut
  {NoStop}%
\bibitem [{\citenamefont {Sheth}\ \emph {et~al.}(2001)\citenamefont {Sheth},
  \citenamefont {Mo},\ and\ \citenamefont {Tormen}}]{Sheth:1999su}%
  \BibitemOpen
  \bibfield  {author} {\bibinfo {author} {\bibfnamefont {R.~K.}\ \bibnamefont
  {Sheth}}, \bibinfo {author} {\bibfnamefont {H.~J.}\ \bibnamefont {Mo}}, \
  and\ \bibinfo {author} {\bibfnamefont {G.}~\bibnamefont {Tormen}},\ }\Doi
  {10.1046/j.1365-8711.2001.04006.x} {\bibfield  {journal} {\bibinfo  {journal}
  {Mon. Not. Roy. Astron. Soc.},\ }\textbf {\bibinfo {volume} {323}},\ \bibinfo
  {pages} {1} (\bibinfo {year} {2001})},\ \Eprint
  {http://arxiv.org/abs/astro-ph/9907024} {arXiv:astro-ph/9907024} \BibitemShut
  {NoStop}%
\bibitem [{\citenamefont {Eisenstein}\ and\ \citenamefont
  {Hu}(1997)}]{Eisenstein:1997jh}%
  \BibitemOpen
  \bibfield  {author} {\bibinfo {author} {\bibfnamefont {D.~J.}\ \bibnamefont
  {Eisenstein}}\ and\ \bibinfo {author} {\bibfnamefont {W.}~\bibnamefont
  {Hu}},\ }\Doi {10.1086/306640} {\bibfield  {journal} {\bibinfo  {journal}
  {Astrophys. J.},\ }\textbf {\bibinfo {volume} {511}},\ \bibinfo {pages} {5}
  (\bibinfo {year} {1997})},\ \Eprint {http://arxiv.org/abs/astro-ph/9710252}
  {arXiv:astro-ph/9710252} \BibitemShut {NoStop}%
\bibitem [{\citenamefont {Navarro}\ \emph {et~al.}(1997)\citenamefont
  {Navarro}, \citenamefont {Frenk},\ and\ \citenamefont
  {White}}]{Navarro:1996gj}%
  \BibitemOpen
  \bibfield  {author} {\bibinfo {author} {\bibfnamefont {J.~F.}\ \bibnamefont
  {Navarro}}, \bibinfo {author} {\bibfnamefont {C.~S.}\ \bibnamefont {Frenk}},
  \ and\ \bibinfo {author} {\bibfnamefont {S.~D.~M.}\ \bibnamefont {White}},\
  }\Doi {10.1086/304888} {\bibfield  {journal} {\bibinfo  {journal} {Astrophys.
  J.},\ }\textbf {\bibinfo {volume} {490}},\ \bibinfo {pages} {493} (\bibinfo
  {year} {1997})},\ \Eprint {http://arxiv.org/abs/astro-ph/9611107}
  {arXiv:astro-ph/9611107} \BibitemShut {NoStop}%
\bibitem [{\citenamefont {White}(2002)}]{White:2002at}%
  \BibitemOpen
  \bibfield  {author} {\bibinfo {author} {\bibfnamefont {M.~J.}\ \bibnamefont
  {White}},\ }\Doi {10.1086/342752} {\bibfield  {journal} {\bibinfo  {journal}
  {Astrophys. J. Suppl.},\ }\textbf {\bibinfo {volume} {143}},\ \bibinfo
  {pages} {241} (\bibinfo {year} {2002})},\ \Eprint
  {http://arxiv.org/abs/astro-ph/0207185} {arXiv:astro-ph/0207185} \BibitemShut
  {NoStop}%
\bibitem [{\citenamefont {Bullock}\ \emph {et~al.}(2001)\citenamefont {Bullock}
  \emph {et~al.}}]{Bullock:1999he}%
  \BibitemOpen
  \bibfield  {author} {\bibinfo {author} {\bibfnamefont {J.~S.}\ \bibnamefont
  {Bullock}} \emph {et~al.},\ }\Doi {10.1046/j.1365-8711.2001.04068.x}
  {\bibfield  {journal} {\bibinfo  {journal} {Mon. Not. Roy. Astron. Soc.},\
  }\textbf {\bibinfo {volume} {321}},\ \bibinfo {pages} {559} (\bibinfo {year}
  {2001})},\ \Eprint {http://arxiv.org/abs/astro-ph/9908159}
  {arXiv:astro-ph/9908159} \BibitemShut {NoStop}%
\bibitem [{\citenamefont {Taylor}\ and\ \citenamefont
  {Navarro}(2001)}]{Taylor:2001bq}%
  \BibitemOpen
  \bibfield  {author} {\bibinfo {author} {\bibfnamefont {J.~E.}\ \bibnamefont
  {Taylor}}\ and\ \bibinfo {author} {\bibfnamefont {J.~F.}\ \bibnamefont
  {Navarro}},\ }\Doi {10.1086/324031} {\bibfield  {journal} {\bibinfo
  {journal} {Astrophys. J.},\ }\textbf {\bibinfo {volume} {563}},\ \bibinfo
  {pages} {483} (\bibinfo {year} {2001})},\ \Eprint
  {http://arxiv.org/abs/astro-ph/0104002} {arXiv:astro-ph/0104002} \BibitemShut
  {NoStop}%
\bibitem [{\citenamefont {Dehnen}\ and\ \citenamefont
  {McLaughlin}(2005)}]{Dehnen:2005cu}%
  \BibitemOpen
  \bibfield  {author} {\bibinfo {author} {\bibfnamefont {W.}~\bibnamefont
  {Dehnen}}\ and\ \bibinfo {author} {\bibfnamefont {D.}~\bibnamefont
  {McLaughlin}},\ }\Doi {10.1111/j.1365-2966.2005.09510.x} {\bibfield
  {journal} {\bibinfo  {journal} {Mon. Not. Roy. Astron. Soc.},\ }\textbf
  {\bibinfo {volume} {363}},\ \bibinfo {pages} {1057} (\bibinfo {year}
  {2005})},\ \Eprint {http://arxiv.org/abs/astro-ph/0506528}
  {arXiv:astro-ph/0506528} \BibitemShut {NoStop}%
\bibitem [{\citenamefont {Kazantzidis}\ \emph {et~al.}(2004)\citenamefont
  {Kazantzidis}, \citenamefont {Magorrian},\ and\ \citenamefont
  {Moore}}]{Kazantzidis:2003im}%
  \BibitemOpen
  \bibfield  {author} {\bibinfo {author} {\bibfnamefont {S.}~\bibnamefont
  {Kazantzidis}}, \bibinfo {author} {\bibfnamefont {J.}~\bibnamefont
  {Magorrian}}, \ and\ \bibinfo {author} {\bibfnamefont {B.}~\bibnamefont
  {Moore}},\ }\Doi {10.1086/380192} {\bibfield  {journal} {\bibinfo  {journal}
  {Astrophys. J.},\ }\textbf {\bibinfo {volume} {601}},\ \bibinfo {pages} {37}
  (\bibinfo {year} {2004})},\ \Eprint {http://arxiv.org/abs/astro-ph/0309517}
  {arXiv:astro-ph/0309517} \BibitemShut {NoStop}%
\bibitem [{\citenamefont {Wojtak}\ \emph {et~al.}(2005)\citenamefont {Wojtak},
  \citenamefont {Lokas}, \citenamefont {Gottloeber},\ and\ \citenamefont
  {Mamon}}]{Wojtak:2005fe}%
  \BibitemOpen
  \bibfield  {author} {\bibinfo {author} {\bibfnamefont {R.}~\bibnamefont
  {Wojtak}}, \bibinfo {author} {\bibfnamefont {E.~L.}\ \bibnamefont {Lokas}},
  \bibinfo {author} {\bibfnamefont {S.}~\bibnamefont {Gottloeber}}, \ and\
  \bibinfo {author} {\bibfnamefont {G.~A.}\ \bibnamefont {Mamon}},\ }\href@noop
  {} {\bibfield  {journal} {\bibinfo  {journal} {Mon. Not. Roy. Astron. Soc.},\
  }\textbf {\bibinfo {volume} {361}},\ \bibinfo {pages} {L1} (\bibinfo {year}
  {2005})},\ \Eprint {http://arxiv.org/abs/astro-ph/0503391}
  {arXiv:astro-ph/0503391} \BibitemShut {NoStop}%
\bibitem [{\citenamefont {Hansen}\ \emph {et~al.}(2006)\citenamefont {Hansen},
  \citenamefont {Moore}, \citenamefont {Zemp},\ and\ \citenamefont
  {Stadel}}]{Hansen:2005yj}%
  \BibitemOpen
  \bibfield  {author} {\bibinfo {author} {\bibfnamefont {S.~H.}\ \bibnamefont
  {Hansen}}, \bibinfo {author} {\bibfnamefont {B.}~\bibnamefont {Moore}},
  \bibinfo {author} {\bibfnamefont {M.}~\bibnamefont {Zemp}}, \ and\ \bibinfo
  {author} {\bibfnamefont {J.}~\bibnamefont {Stadel}},\ }\Doi
  {10.1088/1475-7516/2006/01/014} {\bibfield  {journal} {\bibinfo  {journal}
  {JCAP},\ }\textbf {\bibinfo {volume} {0601}},\ \bibinfo {pages} {014}
  (\bibinfo {year} {2006})},\ \Eprint {http://arxiv.org/abs/astro-ph/0505420}
  {arXiv:astro-ph/0505420} \BibitemShut {NoStop}%
\bibitem [{\citenamefont {Kuhlen}\ \emph {et~al.}(2010)\citenamefont {Kuhlen}
  \emph {et~al.}}]{Kuhlen:2009vh}%
  \BibitemOpen
  \bibfield  {author} {\bibinfo {author} {\bibfnamefont {M.}~\bibnamefont
  {Kuhlen}} \emph {et~al.},\ }\Doi {10.1088/1475-7516/2010/02/030} {\bibfield
  {journal} {\bibinfo  {journal} {JCAP},\ }\textbf {\bibinfo {volume} {1002}},\
  \bibinfo {pages} {030} (\bibinfo {year} {2010})},\ \Eprint
  {http://arxiv.org/abs/0912.2358} {arXiv:0912.2358 [astro-ph.GA]} \BibitemShut
  {NoStop}%
\bibitem [{\citenamefont {Piessens}\ \emph {et~al.}(1983)\citenamefont
  {Piessens}, \citenamefont {de~Doncker-Kapenga},\ and\ \citenamefont
  {Ueberhuber}}]{Piessens:1983}%
  \BibitemOpen
  \bibfield  {author} {\bibinfo {author} {\bibfnamefont {R.}~\bibnamefont
  {Piessens}}, \bibinfo {author} {\bibfnamefont {E.}~\bibnamefont
  {de~Doncker-Kapenga}}, \ and\ \bibinfo {author} {\bibfnamefont {C.~W.}\
  \bibnamefont {Ueberhuber}},\ }\href@noop {} {\emph {\bibinfo {title}
  {{Quadpack: a subroutine package for automatic integration}}}}\ (\bibinfo
  {publisher} {Springer-Verlag},\ \bibinfo {year} {1983})\BibitemShut {NoStop}%
\bibitem [{\citenamefont {Press}\ \emph {et~al.}(2007)\citenamefont {Press},
  \citenamefont {Teukolsky}, \citenamefont {Vetterling},\ and\ \citenamefont
  {Flannery}}]{Press:2007}%
  \BibitemOpen
  \bibfield  {author} {\bibinfo {author} {\bibfnamefont {W.~H.}\ \bibnamefont
  {Press}}, \bibinfo {author} {\bibfnamefont {S.~A.}\ \bibnamefont
  {Teukolsky}}, \bibinfo {author} {\bibfnamefont {W.~T.}\ \bibnamefont
  {Vetterling}}, \ and\ \bibinfo {author} {\bibfnamefont {B.~P.}\ \bibnamefont
  {Flannery}},\ }\enquote {\bibinfo {title} {Numerical recipes: The art of
  scientific computing},}\ \ (\bibinfo  {publisher} {Cambridge University
  Press},\ \bibinfo {year} {2007})\ Chap.\ \bibinfo {chapter} {6.8},\ \bibinfo
  {edition} {3rd}\ ed.\BibitemShut {Stop}%
\end{thebibliography}%

\end{document}